%% file: AZ_fit.tex
\documentclass[fleqn,usenatbib,useAMS,usegraphicx]{mn2e}

\usepackage{amsmath}
\usepackage{mathrsfs}
\usepackage{txfonts}
\usepackage{cancel}

\usepackage{pdflscape}

\usepackage[OT2,OT1]{fontenc}
\newcommand\cyr{
\renewcommand\rmdefault{wncyr}
\renewcommand\sfdefault{wncyss}
\renewcommand\encodingdefault{OT2}
\normalfont\selectfont}
\DeclareTextFontCommand{\textcyr}{\cyr}

\DeclareMathOperator{\arsinh}{arsinh}

\DeclareMathOperator{\erf}{erf}
\DeclareMathOperator{\erfc}{erfc}

\providecommand\abs[1]{\left\lvert{#1}\right\rvert}

\renewcommand\rho\varrho
\renewcommand\pi\upi
\renewcommand\partial\upartial
\renewcommand\le\oldleq
\renewcommand\ge\oldgeq

\newcommand\rmd{\mathrm{d}}
\newcommand\rme{\mathrm{e}}
\newcommand\Beta{\mathrm{B}}

\defcitealias{NFW}{NFW}

\title[density profile fitting functions]
{Fitting functions for dark matter density profiles}
\author[An \& Zhao]
{J. An$^1$\thanks{E-mail:~jinan@nao.cas.cn~(JA),
hz4@st-andrews.ac.uk~(HZ)} and H. Zhao$^2$\footnotemark[1]
\\$^1$ National Astronomical Observatories, Chinese Academy of Sciences,
A20 Datun Road, Chaoyang District, Beijing 100012, PR~China;
\\$^2$ School of Physics and Astronomy,
University of St Andrews, North Haugh,
St Andrews, KY16 9SS, UK.}

\pagerange{\pageref{firstpage}--\pageref{lastpage}}
\volume{000}\pubyear{2012}

\begin{document}
\label{firstpage}
\maketitle

\begin{abstract}\noindent
We present a unified parameterization of the fitting functions suitable
for density profiles of dark matter haloes or elliptical galaxies.
A notable feature is that the classical Einasto profile
appears naturally as the continuous limiting case of the cored subfamily
amongst the double power-law profiles of \citeauthor{Zh96}. Based on this,
we also argue that there is basically no \emph{qualitative} difference
between halo models well-fitted by the Einasto profile and
the standard \citetalias{NFW} model.
This may even be the case quantitatively
unless the resolutions of simulations and the precisions of fittings are
sufficiently high to make meaningful distinction possible.
\end{abstract}

\section{Introduction}

Although it is often claimed that the density profiles of dark matter
haloes found in simulations are well described by simple double power-law
functions of radii \citep*[see e.g.,][henceforth NFW]{NFW}, these are
usually based on the fitting with limited radial coverage.
A possible manifestation of this limitation is that these functions,
when they are extrapolated to the centre, typically possess a singularity,
which is generally unphysical. Real dynamical systems
almost certainly have finite phase-space densities,
finite number densities, and finite escape speeds even at the centre,
as argued by fundamental physics:
(1) any halo with fermions are limited in phase-space density
by the Pauli exclusion principle;
(2) any annihilating dark matter particle is limited
in the number density by the annihilation cross section;
(3) any stellar system is limited by its relaxation time; and
(4) the escape speed is limited at least by the speed of light.
To ameliorate the singularity whilst introducing the fewest
extra parameters is to make the density profile more flexible
in its parameterization.

In fact, the results from more recent high resolution simulations
\citep[see e.g.,][]{Na08} appear to suggest that a different class
of fitting formula such as that of \citet{Ei65,Ei69} introduced
earlier for spherical stellar systems may be more close to the `reality'.
The Einasto profiles by construction have finite densities,
but they are less user friendly given their exponential behaviours.
This motivates us to look for smooth transitions from double power laws
to the Einasto models. Although they seem to be qualitatively
different at first glance, the fitting formulae for the double power laws
and the Einasto profiles are indeed closely related.
For example, the Gaussian and the Plummer profile are limiting cases
of the so-called beta profile for hot X-ray gas haloes,
$\propto (1+r^2/a^2)^{-(\beta+1/2)}$ for
$\beta\rightarrow\infty$ or $\beta=2$ respectively.
This hints to us that it may also be possible to connect
the entire family of double power laws and the Einasto profiles.
We shall show that this is achieved by simply 
redefining parameters of the double power-law models.
This is also analogous to relating an isothermal sphere
to polytropic models as in the infinite-index power law reducing
to an exponential.

We argue that this connection is more than a mere mathematical
trick, for it suggests that a cored double power-law profile with a large
outer power-index value and the Einasto profile should be
qualitatively indistinguishable
if the fitting is mostly weighted near and around the central regions.
Moreover, the resolution limit indicates that the fittings to dark
halo profiles, be they numerical simulations or some observational proxies,
are mostly concerned with the behaviour around the `scale radius', that is,
corresponding to the transition regions in the model. Hence, combined with
the uncertainties in the fittings, it is expected that most cusped double
power-law profiles with a moderate outer slope, such as the \citetalias{NFW}
profile and the \citet{He90} model, may not be well-distinguished from
some cored profiles with an extreme value of the outer slope or even
exponential fall-off if the variations of the logarithmic slope in the
latter models are made to be sufficiently slower than the former. Here we
shall show that, in the context of our extended family of models,
this is indeed the case.

Despite widespread use of the double power-law profile, this possibility
has not been investigated in detail, for in practice most focus on
the double power law has been limited to the two limiting power indices
and scarce attention has been paid to the parameter
controlling the sharpness of the transition. The connection of the
double power-law profiles to the \citeauthor{Ei65} profile, for which the
variation of the shape parameter has received more attention,
thus also highlights the role of the corresponding parameter
in the double power law, and subsequently the above-mentioned
difficulty of distinguishing fitting functions.

\begin{figure}
\includegraphics[width=\hsize]{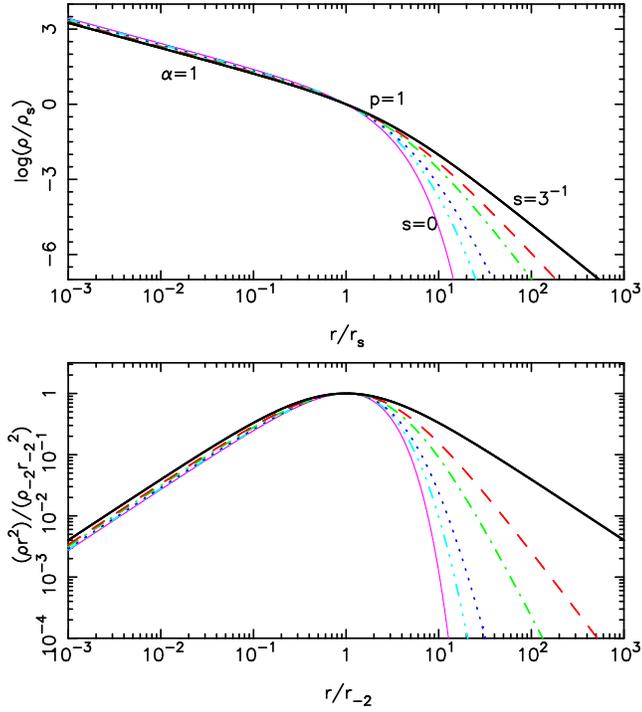}
\caption{\label{fig:den1}
Density profiles of selected members of the family. All models drawn have
$\alpha=1$ and $p=1$ but the values of the parameter $s$ vary with
$\delta=3,4,5,9,15$ and $s=0$, where $\delta=s^{-1}$. The models with
$(\alpha,\delta,p)=(1,3,1)$ and $(\alpha,\delta,p)=(1,4,1)$ correspond to
the so-called \citetalias{NFW} profile and \citeauthor{He90} model,
respectively. In the top panel, the plots are normalized for a fixed
$\rho_\mathrm{s}$ and $r_\mathrm s$ whilst they are for $\rho_{-2}$ and
$r_{-2}$ in the bottom panel. In addition, in the bottom panel, they are
further scaled by $r^{-2}$ so that the horizontal tangent in the figure
indicates that local power index of ${-2}$.}
\end{figure}
\begin{figure}
\includegraphics[width=\hsize]{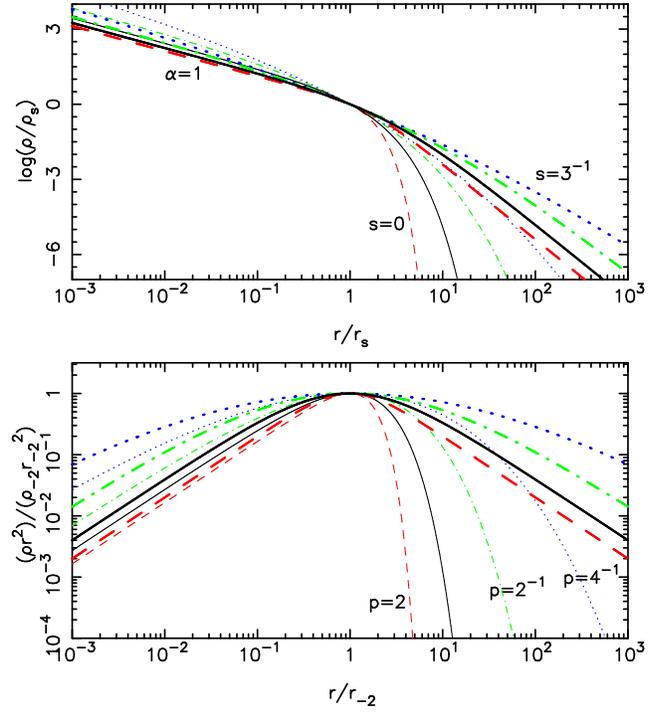}
\caption{\label{fig:den2}
Same as Fig.~\ref{fig:den1} but with different members of the family with
varying $p$. Again all models shown have $r^{-1}$ cusp ($\alpha=1$) but
only those with two different outer slopes, $\delta=3$ (drawn in thick
lines) and an exponential fall-off ($s=0$; thin lines) are chosen. Varying
$p$ values are indicated by different line types: dashed ($p=2$), solid
($p=1$), dot-dashed ($p=\frac12$), and dotted ($p=\frac14$).}
\end{figure}

\section{fitting formulae for density profiles}

\subsection{The Double power law}

A widely-used fitting formula for the density profiles of the dark
haloes found in cosmological simulations is of the form of a `broken' or
double power-law profile \citep{He90,Zh96}:
\begin{equation}\label{eq:p2}
\rho(r)\propto\frac{a^\delta}{r^\alpha(a^p+r^p)^{(\delta-\alpha)/p}}
\,;\qquad
\frac{\rmd\log\rho}{\rmd\log r}
=-\frac{\alpha a^p+\delta r^p}{a^p+r^p}.
\end{equation}
Here, the parameters $\alpha<3$ and $\delta$ correspond to the power indices
of the density profile at the centre and towards the asymptotic limit.
By considering only such profiles with vanishing density at infinity, with
possibility of escape (i.e., the potential at infinity can be set to a
finite value), or with a finite total mass, allowed values of the
parameter $\delta$ may be restricted to be $\delta>0$, $\delta>2$, or
$\delta>3$. The parameter $p>0$ controls the width of the transition region
between the two limiting behaviours; that is, the transition becomes sharper
as $p$ gets larger. Finally, the scale length $a$ in equation (\ref{eq:p2})
specifies the radius at which the local logarithmic density slope is
the same as the arithmetic mean of the two limiting values.

The well-known special cases of these models include:
the \citet{Sc83}--\citet{Pl11} sphere
($\alpha=0$, $\delta=5$, $p=2$: an index-5 polytrope);
the \citet{Ja83} model ($\alpha=2$, $\delta=4$, $p=1$);
the \citet{He90} model ($\alpha=1$, $\delta=4$, $p=1$);
the \citetalias{NFW}
profile ($\alpha=1$, $\delta=3$, $p=1$); and
the isotropic analytic solution of \citet{Au05} and \citet{DM05}
($\alpha=\frac79$, $\delta=\frac{31}9$, $p=\frac49$).
Also notable are the families of models:
the $\gamma$-sphere studied by \citet{De93} and \citet{Tr94}
($\delta=4$, $p=1$);
the generalized NFW profiles of \citet{NFW2,NFW3} ($\delta=3$, $p=1$);
the $\beta$-sphere of \citet{Zh96}
($\alpha=1$, $p=1$: which corresponds to alternative
generalized NFW profiles studied by \citealt{EA06});
the hypervirial family of \citet{EA05}
($\alpha=2-p$, $\delta=p+3$: which was originally introduced
as the generalized isochronous model by \citealt{Ve79});
and the phase-space power-law solutions of \citet{DM05}
($\alpha=1-\frac p2+\beta_0$, $\delta=3+p$: where $\beta_0$ is the
anisotropy parameter at the centre).

\subsection{The Einasto profile}

\citet{Na04} introduced a different class of fitting
formulae for the density profile of dark haloes, namely,
\begin{equation}\label{eq:exp}
\rho\propto\exp\!\left\lgroup-\eta\,\frac{r^p}{a^p}\right\rgroup
\,;\qquad
\frac{\rmd\log\rho}{\rmd\log r}=-p\eta\,\frac{r^p}{a^p}.
\end{equation}
Here $p$ is the parameter controlling how rapidly the logarithmic density
slope varies. Note the $p=2$ case corresponds to the isotropic Gaussian.
The choices of the scale length $a$ and the constant $\eta$ are not
independent but the physical definition of one specifies the other.
For example, if $a=r_{-2}$ where $r_{-2}$ is the radius
at which the logarithmic density slope is `$-2$', then $\eta=\frac2p$.
This family is usually referred to as the Einasto profile
after \citet{Ei65,Ei69}. It also has the same form
as the fitting function proposed by \citet{Se68} and \citet{dV48}
(the latter is a particular case of the former),
which have been used to fit the surface brightness profile
of elliptical (or spheroidal components of) galaxies \citep{Me05}.

\begin{figure}
\includegraphics[width=.95\hsize]{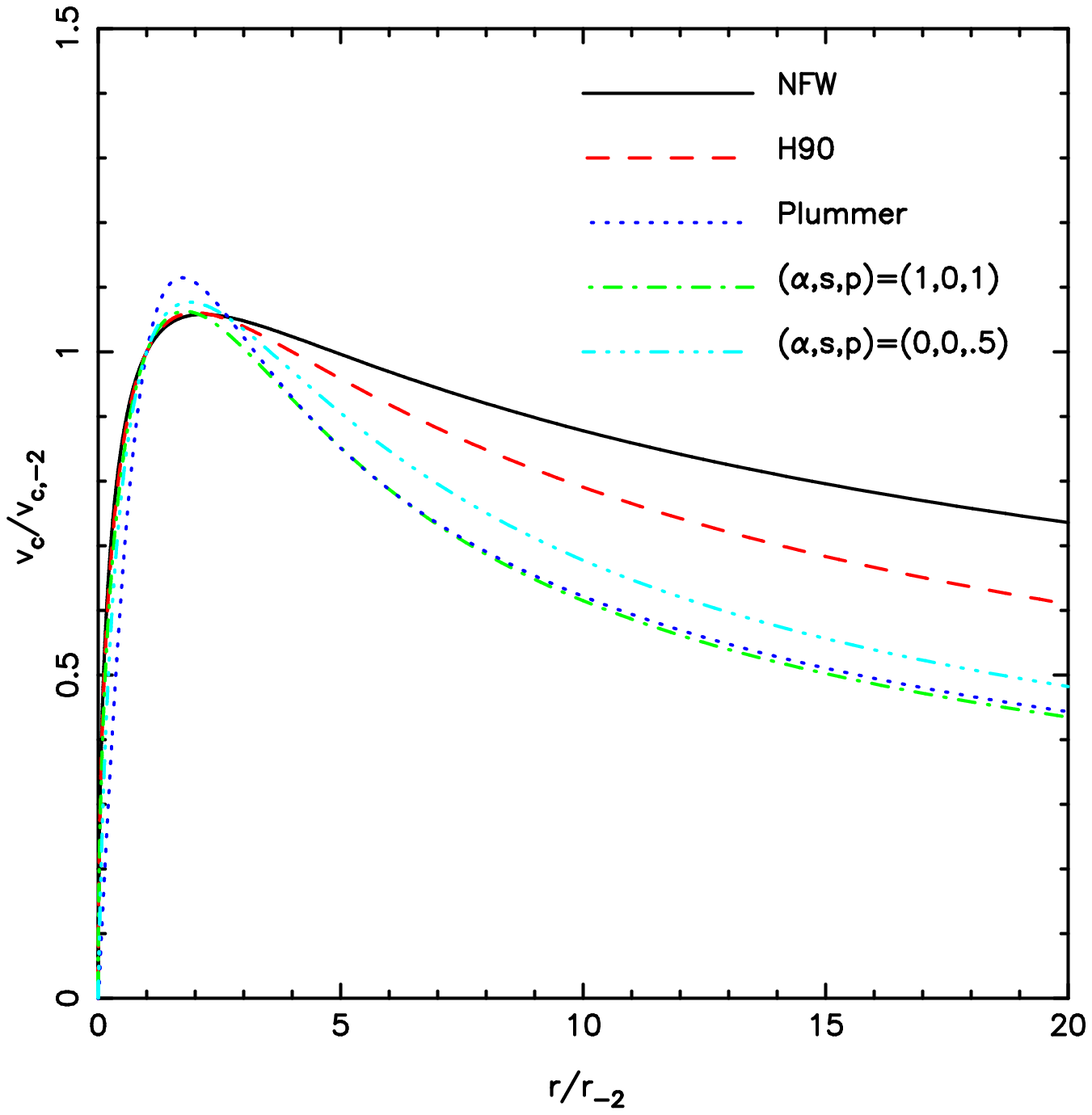}
\caption{\label{fig:circ}
Rotation curve for select models;
\citetalias{NFW} profile ($\alpha=1$, $\delta=3$, $p=1$; solid lines),
the \citeauthor{He90} model ($\alpha=2$, $\delta=4$, $p=1$; dashed lines),
\citeauthor{Sc83}-\citeauthor{Pl11} sphere ($\alpha=0$, $\delta=5$, $p=2$;
dotted lines), $r^{-1}$ cusped exponential fall-off ($\alpha=1$, $s=0$,
$p=1$; dot-dashed lines), and \citeauthor{Se68}-\citeauthor{Ei65} profile
with $p=\frac12$ ($\alpha=0$, $s=0$, $p=\frac12$).
The plots are normalized to have the same circular speed at $r=r_{-2}$.
Among the models shown, only the \citetalias{NFW} profile
has an infinite total mass.
All remaining models have a finite total mass and so their circular speed
speeds fall off as $1/\sqrt r$.}
\end{figure}

\section{An Extension of The Double Power Law}

Functions used to fit the logarithmic density slope in
equations (\ref{eq:p2}) and (\ref{eq:exp}) are first-order
rational functions of $r^p$. Here we consider a slight
variant version of equation (\ref{eq:p2}) given by
\begin{equation}\label{eq:pro}
\frac{\rmd\log\rho}{\rmd\log r}
=-\frac{\alpha r_\mathrm s^p+r^p}{r_\mathrm s^p+sr^p}.
\end{equation}
This is actually the most general first-order monomial rational function
of $r^p$ encompassing both equations (\ref{eq:p2}) and (\ref{eq:exp})
that may be used to fit any physical profile of the logarithm density slope.
Here we assume $r_\mathrm s>0$ without any loss of generality
(the $r_\mathrm s=0$ case corresponds to a pure power law,
which may be described by other parameter combinations).
This reduces to equation (\ref{eq:p2}) through the parameter transformation
$\delta=s^{-1}$ and $a^p=\delta r_\mathrm s^p$,
but it now includes the models with $s=0$ (i.e., $\delta=\infty$)
in places of those with $\delta=0$, which would represent
unphysical asymptotically non-vanishing density profiles.
The $(\alpha,s)=(0,0)$ models correspond to
those of equation (\ref{eq:exp}) with $a^p=p\eta r_\mathrm s^p$.
That is to say, equation (\ref{eq:p2}) with $\alpha=0$ (a cored double
power law) reduces to equation (\ref{eq:exp}) by taking the limit of
$\delta\rightarrow\infty$ and $a\rightarrow\infty$ whilst maintaining
$a^p/\delta=r_\mathrm s^p$ constant
(and redefining $a^p=p\eta r_\mathrm s^p$ afterwards).

In equation (\ref{eq:pro}), the parameter $\alpha<3$ still specifies the
central power index whereas $s>0$ is now the reciprocal of the asymptotic
index. The limits for the finite total mass, the
escapability, and the well-defined integrated density
along any line of sight are given respectively by $0\le s<\frac13$,
$0\le s<\frac12$, and $0\le s<1$. We may introduce a further restriction
$s<\alpha^{-1}$ by limiting ourselves to models whose outer
logarithmic slope is steeper than the inner value. The parameter $p>0$
again controls the steepness of the index variation and the breadth of the
transition region.
On the other hand, although the parameter $r_\mathrm s$ still specifies
the nominal scale length, it lacks any immediate physical interpretation.
By allowing an infinite asymptotic power index,
the arithmetic mean of the limiting indices is not necessarily
defined (note the scale length defined as eq.~\ref{eq:p2} becomes
infinite). An alternative physical scale length may be defined such as the
radius at which the local logarithmic slope is the same as a given fixed
value, e.g.,
\begin{equation}\label{eq:r2}
r_{-n}^p=\frac{n-\alpha}{1-ns}\,r_\mathrm s^p
=\frac{n-\alpha}{\delta-n}\,\delta\,r_\mathrm s^p.
\end{equation}
Here the corresponding logarithmic slope is $(-n)<0$. This is defined
for the models for $\alpha<n$ and $s<n^{-1}$, and simplifies, if $s=0$, to
$r_{-n}^p=(n-\alpha)r_\mathrm s^p$. The most common choice for the
reference logarithmic slope is $(-n)=(-2)$, which corresponds to that of a
singular isothermal sphere.

\subsection{Density profiles}

The actual functional form of the density profile is obtained by
integrating equation (\ref{eq:pro}). If $s>0$, this results in
\begin{subequations}
\begin{equation}\label{eq:den}\begin{split}
\rho(r)&=\rho_\mathrm s\,
(1+s)^{\frac{\delta-\alpha}p}\frac{r_\mathrm s^\alpha}{r^\alpha}
\left(1+\frac{r^p}{\delta\,r_\mathrm s^p}\right)^{-\frac{\delta-\alpha}p}
\\&=\rho_{-2}
\left(\frac{\delta-\alpha}{\delta-2}\right)^{\frac{\delta-\alpha}p}
\left(\frac{r_{-2}}r\right)^\alpha
\left[1+\frac{2-\alpha}{\delta-2}\left(\frac r{r_{-2}}\right)^p
\right]^{-\frac{\delta-\alpha}p},
\end{split}\end{equation}
which reproduces the double power-law profile
(we have used the substitution $\delta=s^{-1}>0$ for clarity).
The integration also introduces an additional constant of integration,
namely, the scale constant.
We have chosen this scale constant in reference to the scale length,
that is, $\rho(r_\mathrm s)=\rho_\mathrm s$ and $\rho(r_{-2})=\rho_{-2}$.
The relation between them is
\begin{equation}
\left(\frac{\rho_{-2}}{\rho_\mathrm s}\right)^p
=\frac{(1-2s)^\delta}{(2-\alpha)^\alpha}\,
\left(\frac{1+\delta}{\delta-\alpha}\right)^{\delta-\alpha}.
\end{equation}
\end{subequations}
We also use
$\bar\rho_\mathrm s\equiv(1+s)^{(\delta-\alpha)/p}\rho_\mathrm s$,
which is useful for compactly writing down many properties of haloes.
This is also related to the proportionality constant of
the usual expression of the double power law in equation (\ref{eq:p2})
such that $C=\bar\rho a^\alpha=\bar\rho_\mathrm sr_\mathrm s^\alpha$
given $a^p=\delta r_\mathrm s^p$.

With $s=0$, the integration leads to the density profile given by
\begin{subequations}
\begin{equation}\label{eq:den0}\begin{split}
\rho(r)&=\rho_\mathrm s\,\rme^{1/p}
\frac{r_\mathrm s^\alpha}{r^\alpha}
\exp\!\left\lgroup-\frac{r^p}{pr_\mathrm s^p}\right\rgroup
\\&=\rho_{-2}\,\left(\frac{r_{-2}}r\right)^\alpha
\exp\!\left\lgroup-\frac{2-\alpha}p
\left[\left(\frac r{r_{-2}}\right)^p-1\right]\right\rgroup,
\end{split}\end{equation}
and the scale constants related by
\begin{equation}
\left(\frac{\rho_{-2}}{\rho_\mathrm s}\right)^p
=\frac{\rme^{\alpha-1}}{(2-\alpha)^\alpha}.
\end{equation}
\end{subequations}
We also define $\bar\rho_\mathrm s\equiv\rme^{1/p}\rho_\mathrm s$,
which is the limit of the prior definition as $s\rightarrow0$.
The transition of models from $s>0$ to $s=0$ is also smooth --
equation (\ref{eq:den0}) is indeed the limit of equation (\ref{eq:den}) as
$s\rightarrow0$ ($\delta\rightarrow\infty$). The profile in equation
(\ref{eq:den0}) is a generalization of the 
Einasto profile of equation (\ref{eq:exp}) allowing a central
density cusp. The classical profile consists of the cored members
($\alpha=0$) of the family with the index $n=p^{-1}$ whilst
a cusped model ($\alpha=-1.85$) has been recently used
to model the Galactic spheroid \citep{Wa12}.

Figures~\ref{fig:den1} and \ref{fig:den2} present the log-log plots
of density profiles for select members of the family.
Here the shape of the profiles is fixed by three of the
parameters $(\alpha,s=\delta^{-1},p)$, whereas varying the
scale length or constant simply translates the graph
horizontally (for the scale length) or vertically (for the scale
constant). Figure~\ref{fig:den1} demonstrates that the
transition of an extreme power law to an exponential fall-off is indeed
natural. The role of parameter $p$ as characterizing the breath of the
transition region is highlighted in Figure~\ref{fig:den2}.

\begin{figure*}
\begin{center}
\includegraphics[width=0.45\hsize]{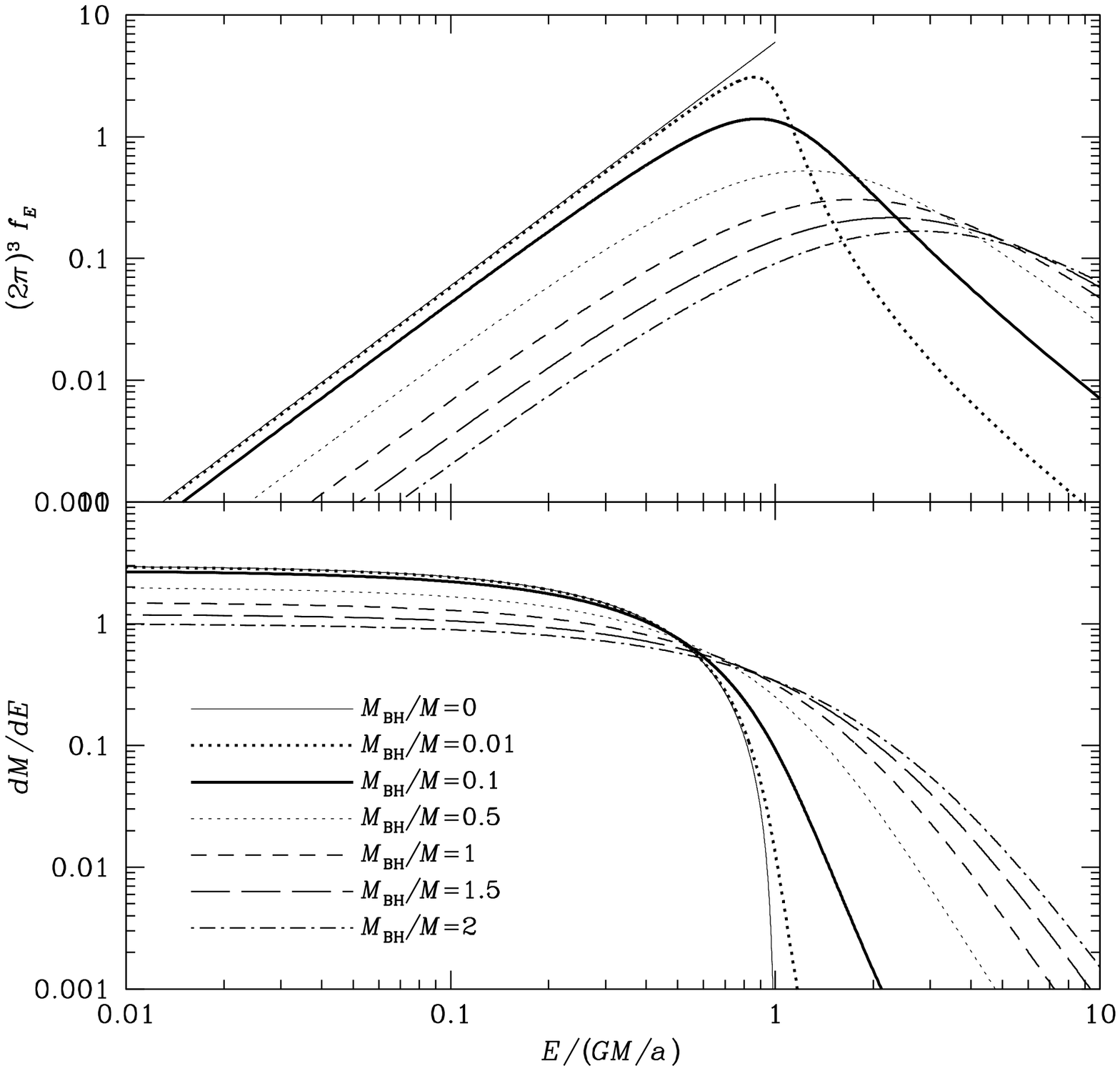}
\includegraphics[width=0.45\hsize]{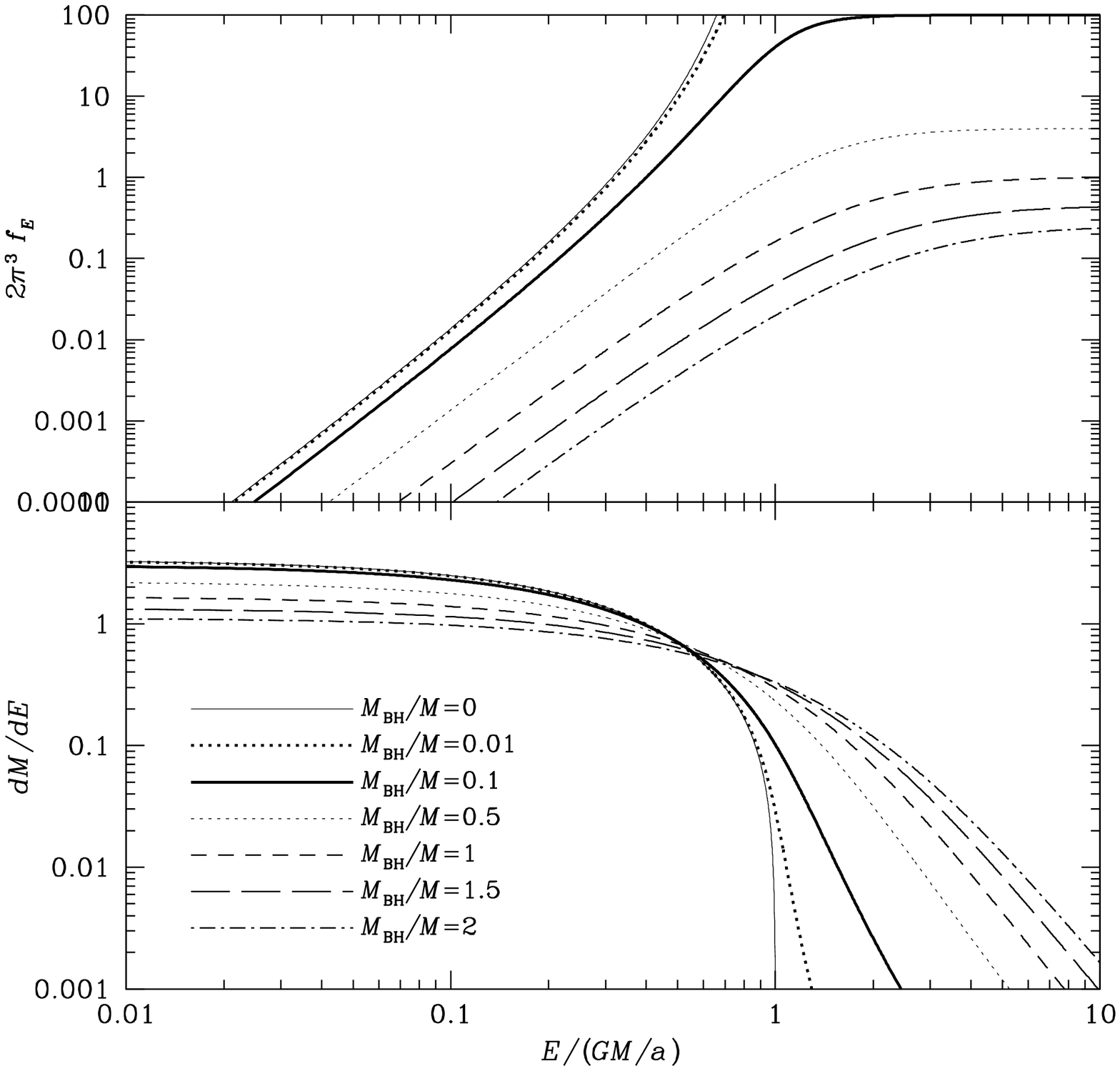}
\end{center}
\caption{\label{fig:hern} Energy part of the distribution function
(upper panel) and the differential energy distribution (lower panel)
of the constant-anisotropy \citeauthor{He90} model with a black hole
for $\beta=\frac12$ (left) and $\beta=-\frac12$ (right):
thin solid lines (no black hole),
thick dotted lines ($M_\bullet=M_\mathrm{tot}/100$),
thick solid lines ($M_\bullet=M_\mathrm{tot}/10$),
thin dotted lines ($M_\bullet=M_\mathrm{tot}/2$),
short-dashed lines ($M_\bullet=M_\mathrm{tot}$),
long-dashed lines ($M_\bullet=3M_\mathrm{tot}/2$),
dot-dashed lines ($M_\bullet=2M_\mathrm{tot}$)
where $M_\mathrm{tot}$ is the total dark halo mass.
All models are normalized to the same $M_\mathrm{tot}$.
Note $0\le\mathcal E\le GM_\mathrm{tot}/a$ when there is no black hole
whereas $\mathcal E\ge0$ otherwise.}
\end{figure*}

\subsection{Basic properties}

Deriving basic properties of these models such as
the enclosed mass profiles $M_r$,
the rotation curves (i.e., the circular speed) $\varv_\mathrm c$,
the potential $\psi$ etc.\ is basically an exercise in integration.
These integrals are trivial to evaluate numerically and
usually result in special functions
of radii no more complicated than beta and gamma functions.
We refer Appendix \ref{app:e} for more systematic studies,
which may be of some peripheral interest.
Here instead we simply provide the rotation curves for some familiar
models and those with exponential fall-off in Figure~\ref{fig:circ}.
Since the total mass is finite for $\delta>3$,
the circular speeds for all plotted models fall off as Keplerian -- i.e.,
$\varv_\mathrm c^2\in\varTheta(r^{-1})$\footnote{$f(x)\in\varTheta[g(x)]$
as $x\rightarrow x_0$
if and only if $\lim_{x\rightarrow x_0}\abs{f/g}$ is non-zero finite.
This is also equivalent to $f\sim Ag$ for a non-zero constant $A$.} --
for large enough radii, except for the \citetalias{NFW} profile,
whose mass profile grows logarithmically
and thus $\varv_\mathrm c^2$ falls off like $r^{-1}\ln r$.
For models with $\delta<3$,
it is obvious that the circular speed would behave like
$\varv_\mathrm c^2\in\varTheta(r^{2-\delta})$ as $r\rightarrow\infty$.
None of the plotted models possesses a cusp
diverging faster or as fast as that of a singular isothermal sphere.
Hence $\varv_\mathrm c^2\in\varTheta(r^{2-\alpha})$ and so
$\varv_\mathrm c^2\rightarrow0$ as $r\rightarrow0$ in Figure~\ref{fig:circ}.
If $\alpha<2<\delta$ like those plotted,
$\varv_\mathrm c$ attains its maximum at the radius corresponding to
the solution of $M_r=4\pi r^3\rho(r)$ (i.e., $\rmd\log M_r/\rmd\log r=1$).
This occurs around $r\approx r_{-2}$ for the models plotted in
Figure~\ref{fig:circ}. Either if $\delta$ is not so much different
from the nominal value of `${-2}$' or if the profile has an extended
transition region (achieved by small $p$),
the variation of $\varv_\mathrm c$ then become sufficiently slow so that
the behaviour of the circular speed can mimic the flat rotation curve
near the region around $r\approx r_{-2}$.

We also find that $M_r\in\varTheta(r^{3-\alpha})$ as $r\rightarrow0$
for $\alpha<3$, which is obvious from the fact that
$\rmd M_r/\rmd r=4\pi r^2\rho\in\varTheta(r^{2-\alpha})$ as $r\rightarrow0$
(assuming the null integration constant). Realistically, any model
with $2<\alpha<3$ cannot be strictly physical extending down
all the way to the centre, for the corresponding
$M_r\in\varTheta(r^{3-\alpha})$ would imply
that there must exist a radius below which $r\le2GM_rc^{-2}$.
This would lead to a gravitational collapse to a singularity and
necessarily a central black hole (or at least the relativistic
effects must be considered for proper understanding).
If the particle dark matter are indeed cold, the origin of the
pressure support is crucial to understand any presence of the density cusp
or core. If the cold dark matter (CDM hereafter) halo consists of
fermions, then the quantum mechanical effects, in particular,
the maximum phase-space density is limited by the mass of the particles
(i.e., ${\rm df}_{\max}\propto m_\mathrm f^4$) and the consequent
degenerate pressure may play a role before the relativistic effects do.
In this regard, it is of special interest to study models with a
finite central density (see Sect.~\ref{sec:core}) which generally have
a finite phase-space density at the centre too, if dark haloes
are limited by the degenerate pressure of fermions.

\subsection{Dynamical properties}

Assessing the dynamical properties associated with these models
typically requires additional assumptions
e.g., regarding the velocity anisotropy etc.
Some of these are straightforward: for instance, under
the ergodic distribution assumption, the unique self-consistent
distribution may be found by means of the \citet{Ed16}
formula although usually the tasks are more involved and in general
the results are somewhat cumbersome. The reader is again referred to
Appendix \ref{app:d} for more systematic studies.
Here we just note that the \citet{He90} model in this regard
is the most special \citep[cf.,][]{BD02}, for a quite a few complete
analytically-tractable dynamical models of the \citeauthor{He90} profile
are known. In Figure~\ref{fig:hern} for example,
we present the distribution function and
the differential energy distribution with the constant anisotropy
given by $\beta=\frac12,-\frac12$ for the dark halo of
the \citeauthor{He90} profile with a central black hole
(see Appendix \ref{app:h} for details).

\begin{figure}
\includegraphics[width=.95\hsize]{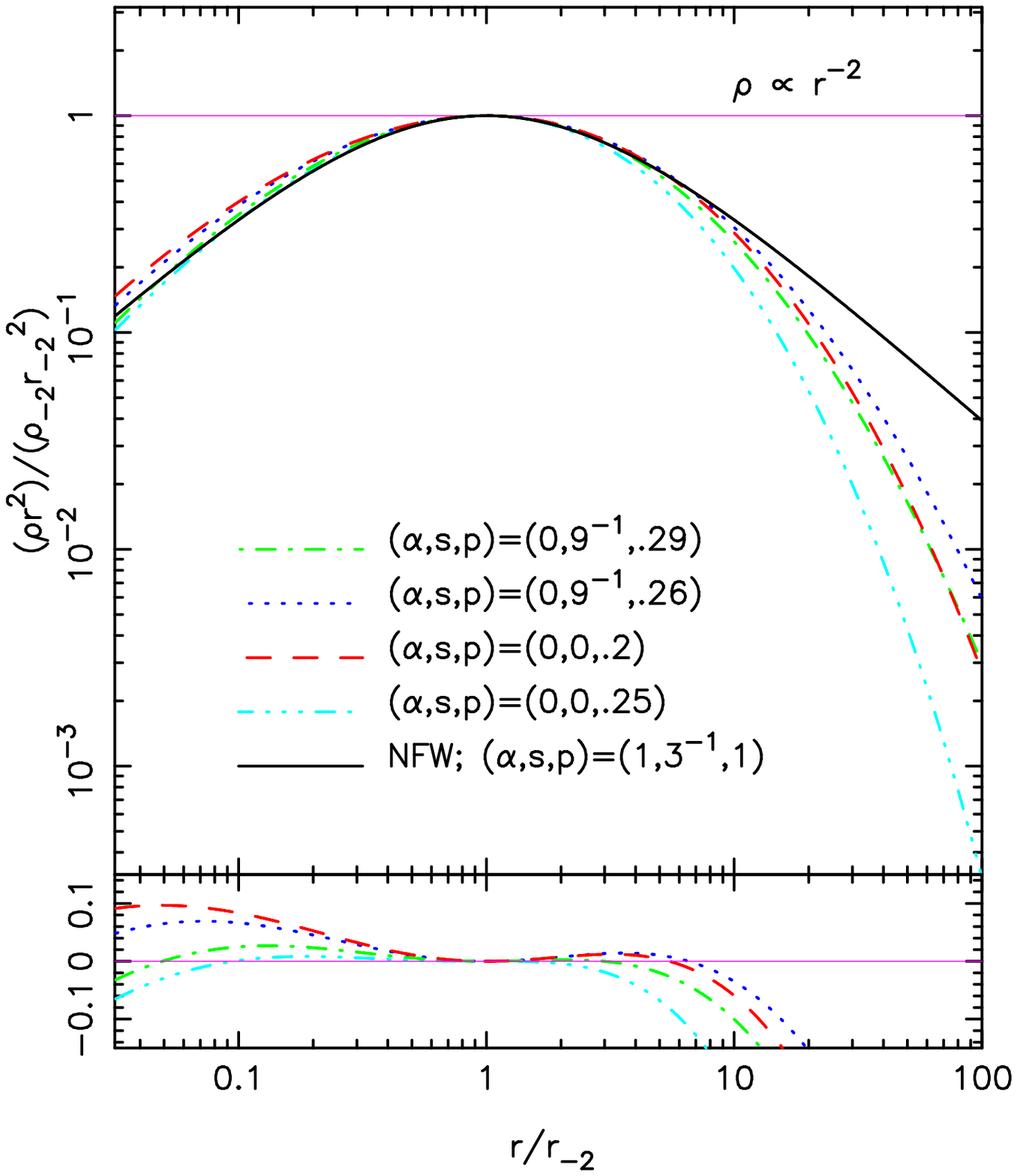}
\includegraphics[width=.95\hsize]{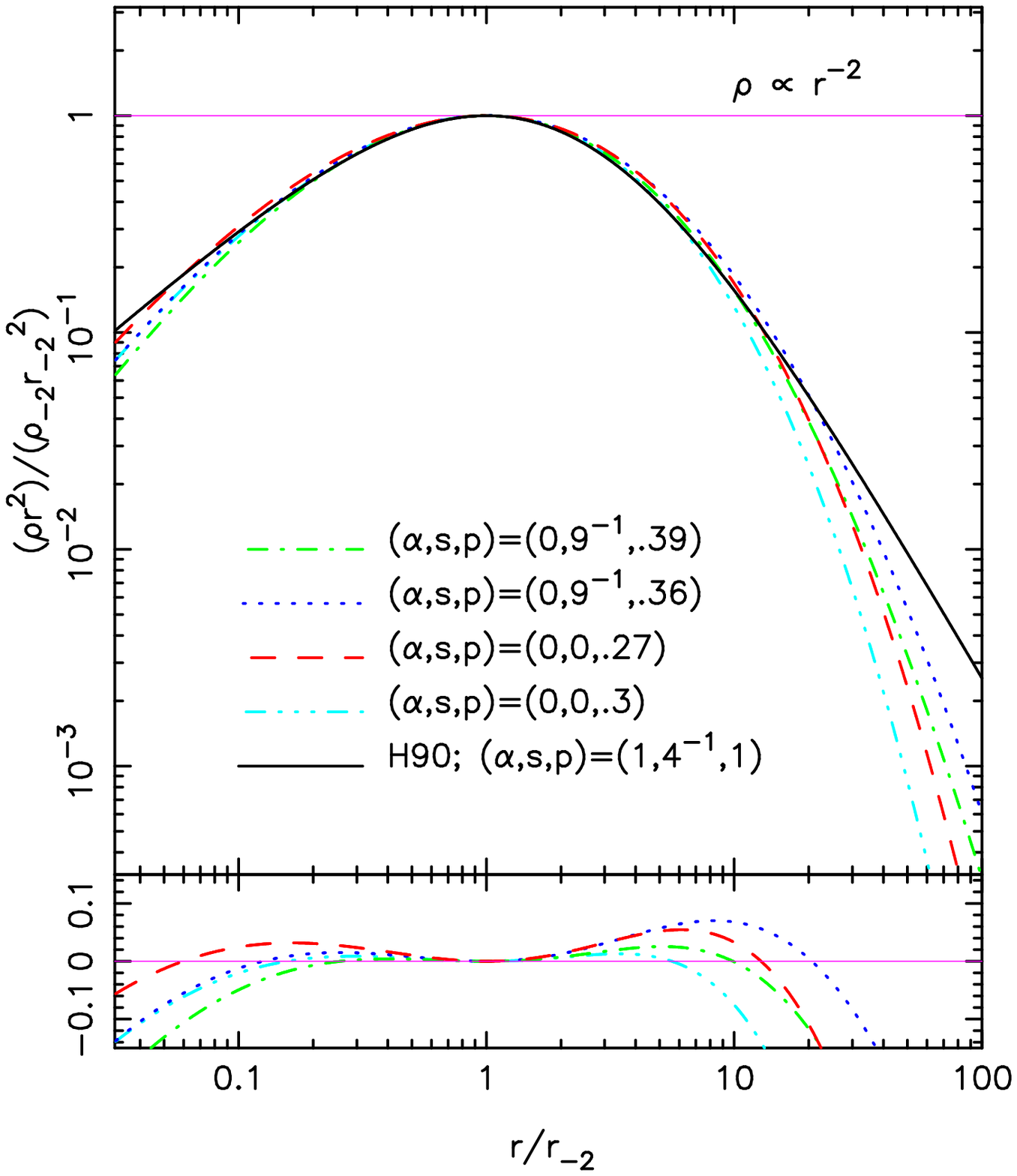}
\caption{\label{fig:ccomp}
Density plots of select cored halo models,
compared to the \citetalias{NFW} profile (top)
and the \citeauthor{He90} model (bottom).}
\end{figure}

\section{The cusp/core problem and densities in projection}

We now turn to comparisons of these models with simulated haloes,
and consider them in the context of the core/cusp problem.

\subsection{A Family of cored profiles}
\label{sec:core}

Whilst earlier N-body simulations seemed to suggest that CDM haloes
follow the universal density profile with a central density cusp,
emerging evidence from recent high resolution simulations indicate
that those findings may have resulted from insufficient
resolving power of earlier simulations. Newer
simulations such as {\sl the Aquaris Project} \citep{Sp08}
that better resolve the central behaviours
show continually varying logarithmic density slopes
(claimed to be well fit by the \citeauthor{Ei65} profile).
We have shown that the \citeauthor{Ei65} profiles are
an extension of the double power law to a member
with a central core and an exponential fall-off.
With a sufficiently slow transition 
and the consequently broad transition zone,
continually varying logarithmic density slopes are
not necessarily inconsistent with the presence of a central core,
given the difficulty of simulating and resolving the central region
of dark haloes. Despite the lack of nominal thermal pressure support,
there are some theoretical arguments, such as the degenerate
pressure of fermion `gas' that argue for a core at the centre of each
CDM halo. All these shepherd us to study the cored members amongst
our model family.

In Figure~\ref{fig:ccomp}, we plot the density profiles of some cored
double power-law models and the \citeauthor{Ei65} profile compared to
the \citetalias{NFW} profile and
the \citeauthor{He90} model. The models are deliberately chosen
so that they closely approximate the latter models
near $r\approx r_{-2}$ except that they are constrained to possess
a core rather than a cusp at the centre.
The finite resolution prevents any information
on the density profile near the very centre from being
available to us. For any practical fitting, thus
there essentially exists an inner cut-off radius below which
the behaviour of the density profile is immaterial.
Furthermore, no dark halo is in isolation in reality and so
the power index for the asymptotic fall-off is
an idealized construct. In fact, the fitting to the density
profile should be cut off before the local halo density approaches
the non-zero background value. However, there is no obvious theoretical
prejudice regarding the location of this truncation radius relative
to the transition zone of the double power-law profile.
That is to say, the outer logarithmic slopes of simulated dark
haloes simply correspond to that at the truncation radius,
rather than the `true' asymptotic slope, and there is no obvious argument
for the former to be close to the latter. With external factors such as
tidal effects and hierarchical environments affecting the outer cut-off,
it is thus possible for the halo to have the local logarithmic slope of
$\approx3$ at truncation as reported by many simulations whilst the `true'
asymptotic slope, if extrapolated, would be steeper (even exponential).
These observations and the uncertainties in the fitting suggest that,
given the fitting is weighted near $r_{-2}$, 
the distinction between the conventional double power law
and a cored profile with a sufficiently broad transition zone
afforded by small $p$ is nontrivial and they may even be degenerate
in some degree.

In practice, the presence of the central density cores versus cusps
in dark haloes may be discriminated by observing $\gammaup$-ray
signals at the Galactic centre due to pair annihilations of CDM particles
\citep[but see also][]{ZS05}. 
In the following (Sect.~\ref{sec:em}), we shall briefly analyze
the behaviour of the so-called astrophysical term $J(\phi)$
\citep[e.g.,][]{La10} in the CDM annihilation signal
for the dark halo density profile given by our model family.
We find that if the dark halo density
profile diverges faster than $r^{-1/2}$ as $r\rightarrow0$,
the astrophysical term also formally diverges as $\phi\rightarrow0$
-- in particular, the $r^{-\alpha}$ cusp ($\alpha>\frac12$) results
in $J\in\varTheta(\phi^{-(2\alpha-1)})$.
Whilst the finite angular resolution
indicates that the observed signal should be finite (unless
the density cusp is steeper than $r^{-3/2}$), the signal from the
$r^{-1}$ cusp of an \citetalias{NFW} halo should be significantly
stronger than that from a cored halo provided that both are normalized
locally in the Solar neighbourhood. For example, \citet{La10} find
that the \citetalias{NFW} halo produces one to two orders of magnitude
stronger annihilation signal towards the Galactic centre than
the \citeauthor{Ei65} model with the resolution ranging from
$\approx3~\mbox{deg}^2$ to $\approx100~\mbox{arcmin}^2$.

\subsection{Surface density profiles}

In terms of the comparison to the real astronomical observations,
it is important to fold the effect of projection into our models.
The most basic among these is the integrated mass density
along the line of sight (i.e., the surface density profile) given by
\begin{equation}
\label{eq:sb}
\Sigma(R)=2\int_R^\infty\frac{\rho(r)\,r\,\rmd r}{\sqrt{r^2-R^2}}.
\end{equation}
This is of use for the lensing convergence $\kappa(R)$ of a halo
or the surface brightness profile $I(R)$ of a `transparent'
constant mass-to-light ratio stellar spheroid, both of which are
proportional to $\Sigma(R)$. If $s\ge1$ ($0\le\delta\le1$), then
equation (\ref{eq:sb}) diverges for any finite $R$ and so $\Sigma(R)$
is not well defined. For the remaining cases that $0\le s<1$ ($\delta>1$),
the result can in principle be evaluated numerically
-- for an arbitrary parameter combination,
the integral in general results in the Fox H-function,
but the particular results are in practice rather `useless'
for the purpose of calculations.

With the cusp/core problem in mind, the important characteristics
of the surface density profiles of note are their limiting behaviours.
The leading terms of $\Sigma(R)$ as $R\rightarrow0$ or $R\rightarrow\infty$
may be obtained by analyzing equation (\ref{eq:sb}) without actually
resolving the integral. The fundamental results in these regards are
\begin{subequations}
\begin{gather}
\frac{\Sigma(0)}{\bar\rho_\mathrm sr_\mathrm s}=
\frac2p\,\Gamma\left(\frac{1-\alpha}p\right)
\times\begin{cases}
\dfrac{\delta^{(1-\alpha)/p}\Gamma\bigl(\frac{\delta-1}p\bigr)}
{\Gamma\bigl(\frac{\delta-\alpha}p\bigr)}&(s>0)\\
p^{(1-\alpha)/p}&(s=0)\end{cases},
\label{eq:lim0}\\\label{eq:limb}
\lim_{R\rightarrow0}\frac{\Sigma(R)}{R\,\rho(R)}
=\Beta\!\left(\frac12,\frac{\alpha-1}2\right)
\,;\quad
\lim_{R\rightarrow\infty}\frac{\Sigma(R)}{R\,\rho(R)}
=\Beta\!\left(\frac12,\frac{\delta-1}2\right),
\end{gather}
\end{subequations}
where $\Gamma(u)$ and $\Beta(u,w)$ are the gamma and beta functions.
Equation (\ref{eq:lim0}) is valid for $\alpha<1$
whereas equations (\ref{eq:lim0}) are finite for $\alpha>1$ and
$\delta>1$, respectively.
That is to say, the surface density at $R=0$ is finite for $0\le\alpha<1$
whilst it is cusped as $R^{-(\alpha-1)}\rightarrow\infty$ near $R=0$
if $\alpha>1$ and it falls off like $R^{-(\delta-1)}$ for $s>0$
($\delta>0$). Please see Appendix \ref{app:p}
\citep[also][sect.~6]{AE09} for more details.

\subsection{The Emission measure}
\label{sec:em}

From the mathematical point of view, the quantity defined such that
\begin{equation}
\mathcal J\equiv\int_{\rm l.o.s.}\!\rho^2\rmd\ell
=2\int_R^\infty\frac{[\rho(r)]^2r\,\rmd r}{\sqrt{r^2-R^2}}
\end{equation}
shares common properties
with the surface density defined in equation (\ref{eq:sb}),
corresponding to the parameter set of $(2\alpha,2\delta)$.
This is related to the emission measure
of the free-free radiation (Bremsstrahlung) and
the so-called astrophysical term in the CDM annihilation signal
from an external dark halo. For the dark halo of the Milky Way,
the astrophysical term defined to be
\begin{equation}
J(\phi)=\begin{cases}{\displaystyle
\int_{r_\phi}^{r_\mathrm v}\!
\frac{[\rho(r)]^2r\,\rmd r}{(r^2-r_\phi^2)^{1/2}}
+\int_{r_\phi}^{r_\odot}\!
\frac{[\rho(r)]^2r\,\rmd r}{(r^2-r_\phi^2)^{1/2}}}
&(0\le\phi\le\frac\pi2)\\{\displaystyle
\int_{r_\odot}^{r_\mathrm v}\!
\frac{[\rho(r)]^2r\,\rmd r}{(r^2-r_\phi^2)^{1/2}}}
&(\frac\pi2\le\phi\le\pi)
\end{cases};
\end{equation}
with $r_\phi=r_\odot\sin\phi$
is more relevant to the actual observable signal.
Here $r_\mathrm v$ is the virial radius of the Galactic dark halo,
$r_\odot$ is the Galactocentric distance of the Sun
and $\phi$ is the angle between the line of sight and the Galactic
centre.

Analogous to $\Sigma(R)$, the general analytical expressions of
$\mathcal J(R)$ or $J(\phi)$ for the three-dimensional density profiles
considered in this paper are expressible using the Fox H-function
whilst their numerical evaluations may be achieved straightforwardly
through the quadrature sum. The limiting behaviours of $\mathcal J$ as
$R\rightarrow0$ are found similarly. In particular,
for a cusped profile $\rho\in\varTheta(r^{-\alpha})$ with $\alpha>0$,
\begin{subequations}
\begin{gather}
\mathcal J\in\begin{cases}
\varTheta(R^{-(2\alpha-1)})&(\alpha>\frac12)\\
\varTheta(\ln R^{-1})&(\alpha=\frac12)\\
\varTheta(1)&(\alpha<\frac12)
\end{cases}\\\nonumber
\abs{\mathcal J(0)-\mathcal J(R)}\sim A'R^{1-2\alpha}
\quad(0<\alpha<\tfrac12),
\end{gather}
whilst we find $\mathcal J\in\varTheta(1)$ and
\begin{equation}
\abs{\mathcal J(0)-\mathcal J(R)}\in\begin{cases}
\varTheta(R^{1+p})&(0<p<1)\\
\varTheta(R^2\ln R^{-1})&(p=1)\\
\varTheta(R^2)&(p>1)
\end{cases}
\end{equation}
\end{subequations}
for a cored profile such that $\abs{\rho(0)-\rho(r)}\sim Ar^p$.
As for $J(\phi)$, we note $J(\phi)\sim\mathcal J(r_\odot\sin\phi)$ and
$\sin\phi\sim\phi$ as $\phi\rightarrow0$. It then follows that
$J(\phi)$ diverges as $\phi^{-(2\alpha-1)}$ or $\ln\phi^{-1}$
for a cusped halo with $\alpha>\frac12$ or $\alpha=\frac12$.
On the other hand, $J(\phi=0)$ should be finite if the Galactic CDM
halo is cored or cusped slower than $r^{-1/2}$. These last cases may
in principle be discriminated by examining $\rmd J/\rmd\phi$, which
formally diverges for a cusped halo as $\phi\rightarrow0$ whilst
it vanishes for a cored halo at the same limit. We leave more detailed
studies of the projected properties of the models for the future.

\small
\section*{acknowledgments}\noindent
Large part of this work by JA was done during the 2008/9 academic year
in the Dark Cosmology Centre,
which is funded by the Danish National Research Foundation
(Danmarks Grundforskningsfond).
He is currently supported by the Chinese Academy of Sciences (CAS)
Fellowships for Young International Scientists (Grant No.: 2009Y2AJ7).
HZ was supported by a visitor grant from the 
National Astronomical Observatories, the Chinese Academy of Sciences (NAOC)
during the final part of the project.

\normalsize
\appendix
\section{Elementary Integral Properties}
\label{app:e}
\subsection{The Enclosed mass}

The mass enclosed in a sphere of radius $r$ is obtained by
\begin{subequations}
\begin{equation}\label{eq:mint}
M_r=4\pi\int_0^r\!\rho(u)\,u^2\rmd u.
\end{equation}
With our models, we should limit $\alpha<3$
since the integrand behaves like
$r^{2-\alpha}$ as $r\rightarrow0$ (if otherwise, this diverges).
This is directly related to the gravitational
acceleration and the circular speed,
\begin{equation}
\left\lVert\bmath g\right\rVert=\frac{\varv_\mathrm c^2}r=\frac{GM_r}{r^2}.
\end{equation}
\end{subequations}

For $\alpha<3$, $p>0$, and $s=\delta^{-1}>0$,
equation (\ref{eq:mint}) results in
\begin{subequations}
\begin{gather}\label{eq:mass}
M_r=\frac{4\pi C_M}p\,
B\left(\frac{x^p}{1+x^p};
\frac{3-\alpha}p,\frac{\delta-3}p\right),
\\\nonumber
C_M=\frac{\bar\rho_\mathrm sr_\mathrm s^3}{s^{(3-\alpha)/p}}
=\rho_{-2}\,r_{-2}^3\left[\frac{(\delta-\alpha)^{\delta-\alpha}}
{(2-\alpha)^{3-\alpha}(\delta-2)^{\delta-3}}
\right]^{1/p}.
\end{gather}
where
\begin{equation}\label{eq:x}
x^p=\frac{r^p}{a^p}
=\frac{r^p}{\delta\,r_\mathrm s^p}
=s\,\frac{r^p}{r_\mathrm s^p}
=\frac{2-\alpha}{\delta-2}\left(\frac r{r_{-2}}\right)^p,
\end{equation}
\end{subequations}
whilst $B(z;u,w)\equiv\Beta_z(u,w)$ is the incomplete beta function.
The beta function is efficiently evaluated using
series or continued fractions,
which converge typically faster than integrating equation (\ref{eq:mint})
via the quadrature sum. Some reduce to expressions involving only
elementary (algebraic and logarithmic or inverse
trigonometric/hyperbolic) functions, which are in general
possible if any of the following is true:
\begin{itemize}
\item $(3-\alpha)/p$ or $(\delta-3)/p$ is a positive integer,
\item $(\delta-\alpha)/p$ is a zero or a negative integer,
\item $(\delta-3)/p$ is an integer and $(3-\alpha)/p$ is a positive
rational, or
\item both $2(3-\alpha)/p$ and $2(\delta-3)/p$ are integers.
\end{itemize}

Equation (\ref{eq:mint}) for $s=0$ on the other hand results in
\begin{subequations}
\begin{gather}\label{eq:mass0}
M_r=\frac{4\pi C_{M,0}}p\,
\gamma\left(\frac{3-\alpha}p;y^p\right),
\\\nonumber
C_{M,0}=\bar\rho_\mathrm sr_\mathrm s^3p^{(3-\alpha)/p}
=\rho_{-2}\,r_{-2}^3\rme^{(2-\alpha)/p}
\left(\frac p{2-\alpha}\right)^{(3-\alpha)/p}.
\end{gather}
where $\alpha<3$, $p>0$ and
\begin{equation}\label{eq:y}
y^p=\frac{r^p}{pr_\mathrm s^p}
=\frac{2-\alpha}p\left(\frac r{r_{-2}}\right)^p.
\end{equation}
\end{subequations}
Here, $\gamma(u;z)$ is the lower incomplete gamma function.
This `simplifies' to elementary (up to exponential) functions
if $(3-\alpha)/p$ is a positive integer. 
It is also recast to formulae using the error function and elementary
functions for half-integer values of $(3-\alpha)/p$.

The behaviour of the integrand, $r^{2-\delta}$ as $r\rightarrow\infty$
of equation (\ref{eq:mint}) indicates that
the finite total mass is defined for $\delta>3$ (including the $s=0$ case).
In particular,
\begin{equation}
M_\mathrm{tot}=\begin{cases}
{\displaystyle\frac{4\pi C_M}p\,
\Beta\!\left(\frac{3-\alpha}p,\frac{\delta-3}p\right)}
&(0<s<\tfrac13)\smallskip\\
{\displaystyle\frac{4\pi C_{M,0}}p\,
\Gamma\!\left(\frac{3-\alpha}p\right)}
&(s=0)\end{cases}
\end{equation}
where $\Beta(u,w)=B(1;u,w)=\Gamma(u)\Gamma(w)/\Gamma(u+w)$ is the
beta function and $\Gamma(u)=\gamma(u;\infty)$ is the gamma function.
If $s\ge\frac13$ ($\delta\le3$) on the other hand,
then the total mass is infinite. Specifically,
the enclosed mass diverges logarithmically
if $\delta=3$ or like $r^{3-\delta}$ if $\delta<3$.

\subsection{The Gravitational potential}

The potential of an isolated system
up to an additive constant is found by integrating
$\bmath\nabla\psi=-\bmath g$. Under spherical symmetry,
\begin{equation}
\frac{\rmd\psi}{\rmd r}=\frac{GM_r}{r^2}
=-\frac\rmd{\rmd r}\left(\frac{GM_r}r\right)+4\pi G\rho r
\end{equation}
where we have used $\rmd M_r/\rmd r=4\pi\rho r^2$.
Then the potential behaves asymptotically
like $r^{-1}$, $r^{-1}\ln r$, $r^{-(\delta-2)}$ and $\ln r$
for $\delta>3$, $\delta=3$, $\delta<3$ but $\delta\ne2$, and $\delta=2$,
respectively. If $\delta>2$, the potential at infinity is not divergent
and any particle with enough kinetic energy can
escape to an unbound state. By contrast, if $\delta\le2$,
then $\psi\rightarrow\infty$ as $r\rightarrow\infty$, and thus
every tracer particle in the model with $s\ge\frac12$
is bound to the system and escape is formally impossible.

If $0\le s<\frac12$ ($\delta>2$),
it is customary to set $\psi(\infty)=0$ and thus
\begin{equation}\label{eq:pot_inf}
\psi(r)-\cancel{\psi_\infty}
=-\frac{GM_r}r-4\pi G\int_r^\infty\!\rho(u)\,u\,\rmd u.
\end{equation}
For $s=\delta^{-1}>0$,
\begin{gather}\label{eq:pot}
\frac\psi{4\pi G}=-\frac{C_\psi}p\left[\frac1x\,
B\left(\frac{x^p}{1+x^p};
\frac{\tilde\alpha_3}p,\frac{\hat\delta_3}p\right)
+B\left(\frac1{1+x^p};
\frac{\hat\delta_2}p,\frac{\tilde\alpha_2}p\right)\right],
\\\nonumber
C_\psi=\frac{C_M}{\delta^{1/p}r_\mathrm s}
=\bar\rho_\mathrm sr_\mathrm s^2\delta^{\tilde\alpha_2/p}
=\rho_{-2}\,r_{-2}^2
\left[\frac{(\delta-\alpha)^{\delta-\alpha}}
{{\tilde\alpha_2}^{\tilde\alpha_2}{\hat\delta_2}^{\hat\delta_2}}
\right]^{1/p},
\end{gather}
whilst, for $s=0$,
\begin{gather}\label{eq:pot0}
\frac\psi{4\pi G}=-\frac{C_{\psi,0}}p
\left[\frac1y\,
\gamma\left(\frac{\tilde\alpha_3}p;y^p\right)
+\varGamma\left(\frac{\tilde\alpha_2}p;y^p\right)\right],
\\\nonumber
C_{\psi,0}
=\frac{C_{M,0}}{p^{1/p}r_\mathrm s}
=\bar\rho_\mathrm sr_\mathrm s^2p^{\tilde\alpha_2/p}
=\rho_{-2}\,r_{-2}^2
\left(\frac\rme{\tilde\alpha_2/p}\right)^{\tilde\alpha_2/p}.
\end{gather}
where $\varGamma(u;z)$ is the upper incomplete gamma function.
We also use short-hand notation,
$\tilde\alpha_n=(n-\alpha)$ and $\hat\delta_n=(\delta-n)$
for an integer $n$.
These again reduce to expressions involving only
elementary functions if $M_r$ is expressible in such a way and the
parameter combinations $\hat\delta_2/p$ and $\tilde\alpha_2/p$ satisfy
criteria similar to those applied for $\hat\delta_3/p$ and
$\tilde\alpha_3/p$.
Some specific examples are found in Appendix~\ref{app:tab}.

The integral in equation (\ref{eq:pot_inf}) for $\delta\le2$ is divergent,
and so equation (\ref{eq:pot}) for such are meaningless
(although they are technically well-defined
unless $\hat\delta_2/p$ is a negative integer).
The potential for $\delta\le2$ however can still be defined
with an alternative zeropoint.
Since $M_r\in\varTheta(r^{3-\alpha})$ as $r\rightarrow0$,
we have $\psi\in\varTheta(r^{2-\alpha})$ if $\alpha\ne2$
or $\varTheta(\ln r^{-1})$ for $\alpha=2$ in the same limit.
Hence, provided that $\alpha<2$
(note that we have generally restricted to be $\alpha<\delta$),
the alternative reference for the potential $\psi(0)=0$ is well-defined
and
\begin{equation}\label{eq:pot_cen}
\psi(r)-\cancel{\psi_0}
=-\frac{GM_r}r+4\pi G\int_0^r\!\rho(u)\,u\,\rmd u.
\end{equation}
For $s>0$, this results in
\begin{equation}
\frac\psi{4\pi G}=\frac{C_\psi}p\left[
B\left(\frac{x^p}{1+x^p};
\frac{\tilde\alpha_2}p,\frac{\hat\delta_2}p\right)
-\frac1x\,
B\left(\frac{x^p}{1+x^p};
\frac{\tilde\alpha_3}p,\frac{\hat\delta_3}p\right)\right]
\end{equation}
If $\alpha<2<\delta$, both this and equation (\ref{eq:pot}) are
well-defined, but they differ by a constant (independent of $r$)
-- note $B(1-z;u,w)+B(z;w,u)=\Beta(u,w)$ for $u,w>0$. 
This difference corresponds to the depth of the central potential well,
$\abs{\psi_\infty-\psi_0}$.
This and $\sqrt{2\abs{\psi_\infty-\psi_0}}$
are also the maximum specific energy and speed
of any bound particle in the system.

For $s=0$, equation (\ref{eq:pot0}) is always (provided that $\alpha<3$)
well-defined. That is to say, since the total mass is finite,
it is always possible to set $\psi(\infty)=0$.
The central potential well depth is
then the limit of equation (\ref{eq:pot0}) as $r\rightarrow0$,
which is finite if $\alpha<2$.
Using this and 
$\gamma(u;z)+\varGamma(u;z)=\Gamma(u)$ for $u>0$,
it is possible to derive an alternative expression for equation
(\ref{eq:pot0}) for $\alpha<2$, that is
\begin{equation}
\frac{\psi_0-\psi}{4\pi G}
=\frac{C_{\psi,0}}p\left[\frac1y\,
\gamma\left(\frac{\tilde\alpha_3}p;y^p\right)
-\gamma\left(\frac{\tilde\alpha_2}p;y^p\right)\right]
\end{equation}
of which the zero point reference is essentially at the centre.

\section{Tractable Dynamical Properties}
\label{app:d}
\subsection{Velocity dispersions}

The velocity dispersion profile of a spherical system is found
by integrating the spherical steady-state Jeans equation,
\begin{equation}\label{eq:sje}
\frac1\rho\frac{\rmd(\rho\sigma_r^2)}{\rmd r}
+\frac{2\sigma_r^2-(\sigma_\theta^2+\sigma_\phi^2)}r
+\frac{GM_r}{r^2}=0.
\end{equation}
Equation (\ref{eq:sje}) may be integrated to be
\begin{equation}\label{eq:sig}
\sigma_r^2(r)=\frac1{\rho(r)}\!
\int_r^\infty\!\rmd s\frac{GM_s}{s^2}\rho(s)
\exp\!\left\lgroup\int_r^s\!\rmd u\frac{2\beta(u)}u\right\rgroup
\end{equation}
given the boundary condition $\lim_{r\rightarrow\infty}\rho\sigma_r^2=0$,
once $\rho(r)$ and $M_r$,
as well as the so-called velocity anisotropy parameter
\begin{equation}
\beta(r)\equiv1-\frac{\sigma_\theta^2+\sigma_\phi^2}{2\sigma_r^2}.
\end{equation}
are specified.
For a monotonically-varying anisotropy parameter,
the parametrization introduced by \citet{BvH07},
\begin{equation}\label{eq:beta}
\beta=\frac{\beta_\infty r^{2q}+\beta_0r_\mathrm a^{2q}}
{r^{2q}+r_\mathrm a^{2q}}
=\frac{\rmd\log\bigl[r^{2\beta_0}
(r^{2q}+r_\mathrm a^{2q})^{(\beta_\infty-\beta_0)/q}\bigr]}{\rmd\log r^2}
\end{equation}
may be useful to represent its behaviour. Then,
equation (\ref{eq:sig}) results in
a simple integral quadrature for $\sigma_r^2(r)$, namely,
\begin{subequations}
\begin{gather}\label{eq:sigr0}
\sigma_r^2(r)=\frac1{h(r)}\!
\int_r^\infty\!\frac{GM_u}{u^2}\,h(u)\,\rmd u;
\\\label{eq:haux}
h(r)=r^{2\beta_0}
(r^{2q}+r_\mathrm a^{2q})^{(\beta_\infty-\beta_0)/q}\rho(r).
\end{gather}
\end{subequations}
which can be evaluated numerically if $M_r$ is tabulated.
Since $h\in\varTheta(r^{2\beta_\infty-\delta})$ as $r\rightarrow\infty$,
this is convergent provided
that $1+\beta_\infty<\delta$ \citep[cf.,][see also \citealt{AE06b}]{Ha04}.
Given $\beta\le1$,
equation (\ref{eq:sigr0}) is therefore well defined for $\delta>2$.

\citet{AE06b,AE09} have analyzed the
leading term behaviours of $\sigma_r^2$ in the limit
of $r\rightarrow0$ and $r\rightarrow\infty$.
The complete analytic expressions for arbitrary
parameter combinations are yet possible only with
higher transcendental functions, which are more or less impractical.
For some parameter combinations however, reductions
to simpler functions are available. If $\beta$
is a linear function of the logarithmic density
slope \citep{HM06}, then $r_\mathrm a=a$ and $2q=p$, and so
$h(r)$ in equation (\ref{eq:haux}) also takes the form of
a double power law for $s>0$. If $M_r$ is
additionally reducible to the sum of such factors (e.g., $\hat\delta_3/p$
or $\tilde\alpha_3/p$ is a positive integer), then $\sigma_r^2$
is expressible using the sum of incomplete beta functions.
The simplest example is the $\delta=p+3$ case, for which
\begin{multline}\label{eq:sigr}
\sigma_r^2
=\frac{4\pi GC_\psi}{p\tilde\alpha_3} x^{\alpha-2\beta_0}(1+x^p)^{1-\frac{2(\beta_\infty-\beta_0)-\tilde\alpha_3}p}
\\\times
B\left(\frac1{1+x^p};
\frac{2(2-\beta_\infty)}p+1,\frac{2(1+\beta_0-\alpha)}p\right).
\end{multline}
This includes the $\gamma$-sphere for which $p=1$ and $\delta=4$,
the model studied by \citet{EA05} for which $(\alpha,\delta)=(2-p,p+3)$,
and the analytical solutions of \citet{DM05} for which
$(\alpha,\delta)=(1+\beta_0-\frac p2,p+3)$. Note
$B(z;u,1)=z^u/u$, and so $\sigma_r^2$ for this last case
results in a double power law as the density profile.
The resulting model with
$p=2(2-\beta_0)(\varepsilon-2)/(\varepsilon+6)$ exhibits radial
power-law behaviour, $\rho/\sigma_r^\varepsilon\propto r^{-\eta}$ where
$\eta=\alpha\varepsilon/2+\alpha-\beta_0\varepsilon$
as noted\footnote{The complete specification of the model involves
7 parameters, but the choice of $\beta_\infty$ is independent from
the others and $\delta=p+3$ indicates that specifying either $p$
or $\delta$ also fixes the other. This effectively leaves 5 parameters
and 3 algebraic constraints, and thus specifying two amongst
$(\alpha,p,\beta_0,\epsilon,\eta)$ completely determines this model.}
by \citet{DM05}. Equation (\ref{eq:sigr}) also indicates that
$\sigma_r^2\sim\sigma^2/(3-2\beta_\infty)
\sim\varv_\mathrm c^2/(p+4-2\beta_\infty)$ as $r\rightarrow\infty$
whereas, as $r\rightarrow0$, we find that
$\sigma_r^2\in\varTheta(r^{\alpha-2\beta_0})$,
$\varTheta(r^{2-\alpha}\ln r^{-1})$, and $\varTheta(r^{2-\alpha})$
for $1+\beta_0>\alpha$, $1+\beta_0=\alpha$, and $1+\beta_0<\alpha$
respectively. These are consistent
with behaviours deduced by \citet{AE06b}.

\subsection{Constant-anisotropy distribution functions}

Consider the phase-space distribution given by
\begin{equation}\label{eq:cbdf}
\mathcal F(\mathcal E,L^2)=\begin{cases}
\dfrac{2^{\beta-\frac32}g(\mathcal E)}
      {\pi^{3/2}\Gamma(1-\beta)\,L^{2\beta}}
&(\beta<1)\smallskip\\
\dfrac{\deltaup(L^2)\,g(\mathcal E)}{\sqrt2\pi^{3/2}}
&(\beta=1)\end{cases}
\end{equation}
where $\mathcal E$ and $L$
are the specific binding energy and the magnitude of the specific
angular momentum and $\deltaup(w)$ is the Dirac delta.
This builds a spherical system with the anisotropy parameter
at all radii given by the parameter value $\beta$.
The function $g(\mathcal E)$ is related to
the tracer energy distribution via
\begin{gather}
\frac{\rmd\rho}{\rmd\mathcal E}
=\frac{\abs{\Psi-\mathcal E}^{\frac12-\beta}
\Theta(\Psi-\mathcal E)}{\Gamma\bigl(\frac32-\beta\bigr)\,r^{2\beta}}
\,g(\mathcal E)
\\\nonumber
\frac{\rmd M}{\rmd\mathcal E}
=\frac{N(\mathcal E)\,g(\mathcal E)}{\Gamma\bigl(\frac32-\beta\bigr)}
\,;\quad
N(\mathcal E)=4\pi\!\int_0^{r_\mathcal E}\,
\abs{\Psi-\mathcal E}^{\frac12-\beta}r^{2(1-\beta)}\rmd r,
\end{gather}
where $\Theta(w)$ is the Heaviside unit step function
and $r_\mathcal E$ is the farthest distance to which
a particle with the binding energy $\mathcal E$ can move, that is,
$\Psi(r_\mathcal E)=\mathcal E$. Here, $\rmd\rho/\rmd\mathcal E$
and $\rmd M/\rmd\mathcal E$ are the local and global differential energy
distributions and $N(\mathcal E)$ is the marginalized
density of states. The function $\Psi(r)$ defined to be
\begin{subequations}
\begin{equation}
\Psi(r)\equiv\begin{cases}
\psi(r_\mathrm{out})-\psi(r)&\text{if $r_\mathrm{out}$ is finite}\\
\psi(\infty)-\psi(r)&
\text{if $r_\mathrm{out}=\infty$ and
$\abs{\psi(\infty)}<\infty$}\\
-\psi(r)&\text{if $r_\mathrm{out}=\infty$ and
$\psi(\infty)\rightarrow\infty$}\\
\end{cases}
\end{equation}
is the positive potential function where
$r_\mathrm{out}$ is the boundary radius.
The specific binding energy is given by
$\mathcal E=\Psi-\frac12\varv^2$ where $\varv$ is the speed of the tracer
with its lower bound for bound particles being
\begin{equation}
\underline{\mathcal E}\equiv\begin{cases}
0&\text{for an escapable system}\\
-\infty&\text{for an infinitely-extended inescapable system}
\end{cases}.
\end{equation}
\end{subequations}

The undetermined function $g(\mathcal E)$ is uniquely
specified by the density profile. That is,
integrating equation (\ref{eq:cbdf}) over the accessible velocity space
at a fixed location results in
the augmented density $h(\Psi)=r^{2\beta}\rho$ as a function of $\Psi$.
The function $g(\mathcal E)$ then may be determined by
its inversion, the \citeauthor{Ed16}--\citeauthor{Cu91} formula,
\begin{multline}\label{eq:disint}
g(\mathcal E)
=\frac1{\Gamma(1-\epsilon)}\frac\rmd{\rmd\mathcal E}\!
\int_{\underline{\mathcal E}}^\mathcal E\!
\frac{h^{(\mu)}(\Psi)\,\rmd\Psi}{(\mathcal E-\Psi)^\epsilon}
\\=\begin{cases}{\displaystyle\frac1{\Gamma(1-\epsilon)}\left[
\int_{\underline{\mathcal E}}^\mathcal E\!\!
\frac{h^{(\mu+1)}(\Psi)\,\rmd\Psi}{(\mathcal E-\Psi)^\epsilon}
+\frac{h^{(\mu)}(\underline{\mathcal E})}
      {(\mathcal E-\underline{\mathcal E})^\epsilon}
\right]}&(0<\epsilon<1)\\\
h^{(\mu)}(\mathcal E)&(\epsilon=0)\end{cases}
\end{multline}
where $\mu=\lfloor\frac32-\beta\rfloor$ and $\epsilon=(\frac32-\beta)-\mu$
are the integer floor and the fractional part of $\frac32-\beta$.
Equation (\ref{eq:disint}) indicates that, given the inverse function
$r(\psi)=\psi^{-1}(\psi)$ of $\psi=\psi(r)$ and
the augmented density $h(\Psi)=[r({-\Psi})]^{2\beta}\rho[r({-\Psi})]$,
the distribution function (df) of the form of equation (\ref{eq:cbdf})
is available as an integral quadrature. If $\beta$ is a half-integer
(i.e., $\beta=\frac12,-\frac12,\dotsc$),
the formula results in pure differentiation
\citep{Cu91,EA06}.

\subsubsection{the $(\alpha,\delta)=(2-p,3+p)$ case}

The simplest example is the $(\alpha,\delta)=(2-p,3+p)$ case, for which
\begin{subequations}
\begin{equation}
h(\Psi)=\frac{\bar\rho a^{2\beta}\Phi^{p+3-2\beta}}
{(1-\Phi^p)^{2(1-\beta)/p-1}}
\,;\quad
\Phi=\frac\Psi{\abs{\psi_\infty-\psi_0}}
=\frac{(p+1)\Psi}{4\pi G\bar\rho a^2}.
\end{equation}
Here, $\abs{\psi_\infty-\psi_0}=GM_\mathrm{tot}/a=4\pi G\bar\rho a^2/(1+p)$
is the depth of the central potential (note $0\le\Phi\le1$).
The df is then found with
\begin{equation}\label{eq:hvdf}\begin{split}
g(\mathcal E)&=
\frac{\bar\rho a^{2\beta}}{\abs{\psi_\infty-\psi_0}^{3/2-\beta}}
\tilde\varg\left(\frac{\mathcal E}{\abs{\psi_\infty-\psi_0}}\right),
\\\tilde\varg(\Phi)&=\frac1{\Gamma(1-\epsilon)}\!\int_0^\Phi\!\!
\frac{\rmd Q}{(\Phi-Q)^\epsilon}
\frac{\rmd^{\mu+1}}{\rmd Q^{\mu+1}}\!
\left[\frac{Q^{p+3-2\beta}}{(1-Q^p)^{2(1-\beta)/p-1}}\right].
\end{split}\end{equation}
\end{subequations}
If $2(1-\beta)\ge p>0$, the df is non-negative
for all accessible binding energy interval,
$0\le\mathcal E\le\abs{\psi_\infty-\psi_0}$ \citep[cf.,][]{AE06b}.
Equation (\ref{eq:hvdf}) with arbitrary $p>0$ and $\beta<1$
results in the Fox H-function, which is equivalent
a power series of $\Phi$, namely,
\begin{equation}\label{eq:hvdfs}
\tilde\varg(\Phi)=\Phi^{p+\frac32-\beta}\sum_{k=0}^\infty
\frac{(\beta_p)_k\Gamma(pk+p+4-2\beta)}
{k!\Gamma\bigl(pk+p+\frac52-\beta\bigr)}\,\Phi^{pk},
\end{equation}
where $\beta_p=2(1-\beta)/p-1>0$ and $(a)_k$ is the Pochhammer symbol.
This reduces to a hypergeometric function if $p$ is rational.
The result for the \citeauthor{He90} model was derived by \citet{BD02},
\begin{equation}\label{eq:hern}
\tilde\varg(\Phi)=\frac{\Gamma(5-2\beta)}{\Gamma\bigl(\frac72-\beta\bigr)}\,
\Phi^{\frac52-\beta}
{}_2F_1\Bigl(1-2\beta,5-2\beta;\tfrac72-\beta;\Phi\Bigr),
\end{equation}
whilst for the \citeauthor{Sc83}-\citeauthor{Pl11} sphere, this becomes
\begin{equation}
\tilde\varg(\Phi)=\frac{\Gamma(6-2\beta)}{\Gamma(\frac92-\beta)}\,
\Phi^{\frac72-\beta}
{}_3F_2\Bigl(-\beta,3-\beta,\tfrac72-\beta;
\tfrac{9-2\beta}4,\tfrac{11-2\beta}4;\Phi^2\Bigr).
\end{equation}
If $\beta$ is a half-integer, both are expressible
using only algebraic functions whereas equation (\ref{eq:hern}) with
an integer $\beta$ reduces to an expression involving
up to inverse trigonometric functions, e.g.,
for the isotropic \citeauthor{He90} model,
\begin{equation}
\tilde\varg(\Phi)=\frac1{2\!\!\sqrt\pi}\left[
\frac{3\arcsin\!\!\sqrt\Phi}{(1-\Phi)^\frac52}
-\frac{\!\!\sqrt\Phi\,(1-2\Phi)(3+8\Phi-8\Phi^2)}{(1-\Phi)^2}\right].
\end{equation}

If $2(1-\beta)=p$, equation (\ref{eq:hvdf}) indicates that
$\tilde\varg(\Phi)\propto\Phi^{p+\frac32-\beta}=\Phi^{(3p+1)/2}$.
Equation (\ref{eq:hvdfs}) is still
valid because the series then terminates for $k\ge1$.
The corresponding df,
$\mathcal F(\mathcal E,L^2)\propto L^{p-2}\mathcal E^{(3p+1)/2}$
is the hypervirial model studied by \citet{EA05}.

\subsubsection{the distribution function as a function of $r_\mathcal E$}

Even if an analytic expression for the inverse function of
the potential is not available, equation (\ref{eq:disint})
can be used to write down $g(\mathcal E)$ implicitly
using $r_\mathcal E=\Psi^{-1}(\mathcal E)=\psi^{-1}({-\mathcal E})$.
This is achieved through change of variables for the integral
in equation (\ref{eq:disint}),
\begin{subequations}
\begin{gather}\begin{split}
g(\mathcal E)
&=\frac1{\Gamma(1-\epsilon)}\frac{\rmd^{\mu+1}}{\rmd\mathcal E^{\mu+1}}\!
\int_{r_\mathcal E}^\infty
\frac{GM_r\,\rho(r)\,\rmd r}{r^{2(1-\beta)}[\mathcal E-\Psi(r)]^\epsilon}
\\&=\frac{(-1)^\mu}{\Gamma(1-\epsilon)}
\frac{r_\mathcal E^2}{M_{r_\mathcal E}}
\frac\rmd{\rmd r_\mathcal E}\!\int_\infty^{r_\mathcal E}
\frac{M_r\,\hat\varg_{2\beta}^\mu(r)\,\rmd r}{r^2[\mathcal E-\Psi(r)]^\epsilon}
\end{split};\label{eq:disintr}
\\
\hat\varg_{2\beta}^\mu(r)=
\left(\frac{r^2}{GM_r}\frac\rmd{\rmd r}\right)^\mu
[r^{2\beta}\rho(r)].
\end{gather}
Here $\Psi=-\psi$ and $\rmd\psi/\rmd r=GM_r/r$. In addition,
$\Psi(r_\mathcal E)=\mathcal E$ and so
\begin{equation*}
\frac{\rmd\mathcal E}{\rmd r_\mathcal E}
=-\frac{\rmd\psi}{\rmd r}\biggr|_{r=r_\mathcal E}
=-\frac{GM_{r_\mathcal E}}{r_\mathcal E^2}
\,;\quad
\frac\rmd{\rmd\mathcal E}=-\frac{r_\mathcal E^2}{GM_{r_\mathcal E}}
\frac\rmd{\rmd r_\mathcal E}.
\end{equation*}
\end{subequations}
The integral in equation (\ref{eq:disintr}) for
a positive integer $n=\frac32-\beta$ collapses thanks to
the fundamental theorem of calculus and so
\begin{subequations}\label{eq:ndist}
\begin{equation}
g(\mathcal E)=(-1)^n\hat\varg_{3-2n}^n(r_\mathcal E).
\end{equation}
For $\beta=\frac12$ ($n=1$) and $\beta=-\frac12$ ($n=2$),
this reduces to
\begin{gather*}
-\hat\varg^1_1(r)=
\frac{r^2\rho}{GM_r}\left(-1-\frac{\rmd\log\rho}{\rmd\log r}\right).
\\\hat\varg^2_{-1}(r)=
\frac{r\rho}{(GM_r)^2}\left[\frac{\rmd^2\log\rho}{{\rmd\log r}^2}+
\left(\frac{4\pi r^3\rho}{M_r}-\frac{\rmd\log\rho}{\rmd\log r}\right)
\left(1-\frac{\rmd\log\rho}{\rmd\log r}\right)\right].
\end{gather*}
The df and the differential energy distribution are given by
\begin{gather}
\mathcal F(\mathcal E,L^2)=\frac{g(\mathcal E)}{2\pi^2L}\,;\quad
\frac{\rmd M}{\rmd\mathcal E}=2\pi r_\mathcal E^2g(\mathcal E)
\qquad\text{for $\beta=\tfrac12$,}
\intertext{and}
\mathcal F(\mathcal E,L^2)=\frac L{2\pi^2}g(\mathcal E)\,;
\\\frac{\rmd M}{\rmd\mathcal E}
=4\pi\,g(\mathcal E)\!\int_0^{r_\mathcal E}\!(\Psi-\mathcal E)r^3\rmd r
=\pi\,g(\mathcal E)\!\int_0^{r_\mathcal E}\!GM_rr^2\rmd r
\nonumber\end{gather}
\end{subequations}
for $\beta=-\frac12$. Writing down the corresponding df for our models
using equations (\ref{eq:ndist}) is trivial if tedious.
See \citet{EA06} for simple cases
such as $(p,\delta)=(1,4)$ and $(p,\alpha)=(1,1)$.

\subsubsection{a central black hole}
\label{app:h}

The outlined procedure is valid even if
there exists a central point mass \`a la a black hole.
This does not contribute the local density except at the centre,
and so it is simply incorporated by $M_r=M_{r,\rm halo}+M_\bullet$
and $\psi=\psi_\mathrm{halo}-GM_\bullet r^{-1}$, that is, the mass of
the central black hole $M_\bullet$ is an integration constant.
With non-zero $M_\bullet>0$, the potential $\psi(r)$ is invertible
for the \citeauthor{He90} model and those corresponding to
$(\alpha,\delta,p)=(\frac52,4,1)$ and $(1,\frac52,1)$.
If $\beta=\frac12$, the \citeauthor{He90}
model with a central black hole \citep{Ci96} is built by
\begin{subequations}
\begin{gather}
\mathcal F(\mathcal E,L^2)
=\frac3{4\pi^3aGL}\frac{\varw^2(1-\varw)^2}{\varw^2+\mathscr M};\quad
\frac{\rmd M}{\rmd\mathcal E}=\frac{3a}G\frac{\varw^4}{\varw^2+\mathscr M};
\nonumber\\
\varw=\frac{1-\mathscr E-\mathscr M+
\sqrt{(1-\mathscr E-\mathscr M)^2+4\mathscr M}}2.
\label{eq:yeh}
\end{gather}
For $(\alpha,\delta,p,\beta)=(\frac52,4,1,\frac12)$,
we find \citep{BD04}
\begin{gather}
\mathcal F(\mathcal E,L^2)=\frac3{32\pi^3aGL}
\frac{(1-\varw^2)^2(1+\varw^2)}{\varw(\varw+\mathscr M)};\quad
\frac{\rmd M}{\rmd\mathcal E}
=\frac{3a}{8G}\frac{\varw^3(1+\varw^2)}{\varw+\mathscr M};
\nonumber\\
\varw=\frac{1+\sqrt{(1+\mathscr M)^2+\mathscr{EM}}}
{2+\mathscr E+\mathscr M},
\end{gather}
whilst the model with $(\alpha,\delta,p,\beta)=(1,\frac52,1,\frac12)$
is given by
\begin{gather}
\mathcal F(\mathcal E,L^2)=\frac3{(2\pi)^3aGL}
\frac{z(1-z)^2}{4(1-\!\sqrt z)^2+\!\sqrt z\mathscr M}\,;\\\nonumber
\frac{\rmd M}{\rmd\mathcal E}=\frac{3a}{2G}
\frac{(1-z)^4}
{z[4(1-\!\sqrt z)^2+\!\sqrt z\mathscr M]},
\\\nonumber
\varw=\frac{8(4-\mathscr E)-(16-\mathscr E-\mathscr M)\mathscr M+
8\sqrt{(4-\mathscr E)^2+\mathscr{EM}}}
{(8-\mathscr E-\mathscr M)^2}
\end{gather}
with $z=1-\varw$. For all cases, $\mathscr E$ and $\mathscr M$
are the binding energy and black hole mass normalized such that
\begin{equation*}
\mathscr E=\frac{(3-\alpha)\mathcal E}{4\pi G\bar\rho a^2}\,;\qquad
\mathscr M=\frac{(3-\alpha)M_\bullet}{4\pi\bar\rho a^3}.
\end{equation*}
\end{subequations}
Calculations for $\beta=-\frac12$ are still trivial
albeit messier. We only provide the explicit result
for the \citeauthor{He90} model,
\begin{gather}
\mathcal F(\mathcal E,L^2)=\frac{L\,(1-\varw)^3}{2\pi^3a^2G^2M_\mathrm{tot}}
\frac{\varw^2(3+4\varw+3\varw^2)+\mathscr M(1+3\varw+6\varw^2)}{(\varw^2+\mathscr M)^3}
\nonumber\\
\frac{\rmd M}{\rmd\mathcal E}=\frac a{3G}
\frac{\varw^2(3+4\varw+3\varw^2)+\mathscr M(1+3\varw+6\varw^2)}{(\varw^2+\mathscr M)^3}
\\\nonumber\quad\times
\left[\varw(12-30\varw+22\varw^2-3\varw^3)+\mathscr M\varw^3
+12(1-\varw)^3\ln(1-\varw)\right],
\end{gather}
where $\varw$ is as in equation (\ref{eq:yeh}), and
$\mathscr E=a\mathcal E/(GM_\mathrm{tot})$ and
$\mathscr M=M_\bullet/M_\mathrm{tot}$ where $M_\mathrm{tot}$ is the total halo mass
(sans the black hole).

\section{\boldmath Limiting behaviours of $\lowercase{\Sigma}(R)$}
\label{app:p}

For an arbitrary parameter combination,
equation (\ref{eq:sb}) results in 
\begin{equation}\label{eq:sbf}
\frac{\Sigma(R)}{\bar\rho_\mathrm sr_\mathrm s}
=\biggl(\frac{r_\mathrm s}R\biggr)^{\alpha-1}\times\,
\begin{cases}{\displaystyle
\frac{\sqrt\pi}{\Gamma(\mathfrak d)}\,
{\bf H}^{2,1}_{2,2}\!\left\lgroup 
\frac1\delta\biggl(\frac R{r_\mathrm s}\biggr)^p\
\vrule\begin{array}c
\{1-\mathfrak d,1\},\bigl\{\frac\alpha2,\frac p2\bigr\}\\
\{0,1\},\bigl\{\frac{\alpha-1}2,\frac p2\bigr\}
\end{array}\right\rgroup}
\smallskip\\{\displaystyle
\sqrt\pi\,{\bf H}^{2,0}_{1,2}\!\left\lgroup 
\frac1p\biggl(\frac R{r_\mathrm s}\biggr)^p\
\vrule\begin{array}c
\bigl\{\frac\alpha2,\frac p2\bigr\}\\
\{0,1\},\bigl\{\frac{\alpha-1}2,\frac p2\bigr\}
\end{array}\right\rgroup}
\end{cases}
\end{equation}
where $\mathfrak d=(\delta-\alpha)/p$.
%
%
The first case is for $0<s=\delta^{-1}<1$
whilst the second case corresponds to $s=0$.
For a rational value of $p$, this Fox H-function
reduces to the Me{\ij}er G-function, and subsequently it is possible to
write down using the hypergeometric functions, if desired.
If $p=2$ or $p=1$, this reduces to simpler special functions
or elementary functions for some parameter
combinations. More detailed studies of these types
will be explored elsewhere.

We also find that if $1<\alpha<\delta$, then
\begin{equation}
\frac{\Sigma(R)}{\bar\rho_\mathrm sr_\mathrm s}\simeq
\biggl(\frac{r_\mathrm s}R\biggr)^{\alpha-1}
\left[\frac{\!\sqrt\pi\,\Gamma\bigl(\frac{\alpha-1}2\bigr)}
{\Gamma\bigl(\frac\alpha2\bigr)}+\mathcal O(\varg)\right],
\qquad(R\rightarrow0)
\end{equation}
where the reminder term $\varg(R)$ behaves like
\begin{displaymath}
\varg(R)=\begin{cases}
R^p&(p<\alpha-1)\\R^p\ln R^{-1}&(p=\alpha-1)\\
R^{\alpha-1}&(p>\alpha-1)\end{cases}.
\end{displaymath}
If $\alpha=1<\delta$, then $\Sigma(R)$ diverges logarithmically like
\begin{gather}
\frac{\Sigma(R)}{2\bar\rho_\mathrm sr_\mathrm s}\simeq
\ln\Bigl(\frac{2r_\mathrm s}R\Bigr)-\frac{\gammaup-\mathrm{F}}p+\mathcal O(\varg),
\qquad(R\rightarrow0)
\\\nonumber
\mathrm{F}=\begin{cases}
\ln\delta-\digamma\bigl(\frac{\delta-1}p\bigr)
&(s>0)\\\ln p&(s=0)
\end{cases},
\end{gather}
where $\digamma(z)\equiv\rmd\ln\Gamma(z)/\rmd z$ is the digamma
function and $\gammaup\approx0.5772\dotso$ is the Euler--Mascheroni
constant with the reminder
\begin{displaymath}
\varg(R)=\begin{cases}
R^p&(0<p<2\ \wedge\ p\ne1)\\R^2\ln R^{-1}&(p=1\ \vee\ p=2)\\R^2&(p>2)
\end{cases}.\end{displaymath}
For $0<\alpha<1<\delta$ on the other hand, it can also be shown that
\begin{gather}
\frac{\Sigma(R)-\Sigma(0)}{\bar\rho_\mathrm sr_\mathrm s}\simeq
-\frac{2\!\sqrt\pi\,\Gamma\bigl(\frac{\alpha+1}2\bigr)}
{(1-\alpha)\Gamma\bigl(\frac\alpha2\bigr)}
\biggl(\frac R{r_\mathrm s}\biggr)^{1-\alpha}+\mathcal O(\varg),
\qquad(R\rightarrow0)\\\nonumber
\varg(R)=\begin{cases}
R^{p+1-\alpha}&(p<\alpha+1)\\R^2\ln R^{-1}&(p=\alpha+1)\\
R^2&(p>\alpha+1)\end{cases}.
\end{gather}
In other words, $\Sigma(R)$ for $0<\alpha<1$ decreases outwardly
from the central line of sight like
$\abs{\Sigma(0)-\Sigma(R)}\sim AR^{1-\alpha}$
and possesses a `spiky core', i.e.,
$\rmd\Sigma/\rmd R$ diverges as $R\rightarrow0$. With
$\Sigma(R)$ corresponding to a cored profile with $\alpha=0$, we
find that it exhibits a `flat-topped core' at the centre as
\begin{gather}
\frac{\Sigma(0)-\Sigma(R)}{\bar\rho_\mathrm sr_\mathrm s}\simeq\begin{cases}
\mathrm{G}^{-1}\Gamma\Bigl(1-\frac1p\Bigr)\,\biggl(\dfrac R{r_\mathrm s}\biggr)^2
+\mathcal O(\varg)&(p>1)\\
\biggl[\ln\Bigl(\dfrac{2r_\mathrm s}R\Bigr)-\mathrm{H}+\tfrac12\biggr]
\biggl(\dfrac R{r_\mathrm s}\biggr)^2
+\mathcal O(R^4\ln R^{-1})&(p=1)\smallskip\\
\dfrac{\!\sqrt\pi\,\Gamma\bigl(\frac{1-p}2\bigr)}
{(1+p)\Gamma\bigl(1-\frac p2\bigr)}
\biggl(\dfrac R{r_\mathrm s}\biggr)^{1+p}
+\mathcal O(\varg)&(p<1)\end{cases},
\nonumber\\
\mathrm{G}=\begin{cases}
\dfrac{\delta^{1/p}\Gamma\bigl(\frac\delta p\bigr)}
{\Gamma\bigl(\frac{\delta+1}p\bigr)}&(s>0)\\p^{1/p}&(s=0)
\end{cases}
\,;\quad
\mathrm{H}=\begin{cases}
H_\delta-\ln\delta&(s>0)\\\gammaup&(s=0)
\end{cases},
\end{gather}
where $H_n=\digamma(n+1)+\gammaup$ is the harmonic number.
That is, given the cored 3-d density profile
$\abs{\rho(0)-\rho(r)}\sim Ar^p$ as $r\rightarrow0$, the corresponding
surface density as $R\rightarrow0$ behaves as
$\abs{\Sigma(0)-\Sigma(R)}\sim A'R^{\min(1+p,2)}$ unless $p=1$,
for which it does as $\abs{\Sigma(0)-\Sigma(R)}\sim A'R^2\ln R^{-1}$.
The order of the reminder term is given by
\begin{displaymath}
\varg(R)=\begin{cases}
R^4&(p>3\ \vee\ p=2)\\
R^4\ln R^{-1}&(p=3\ \vee\ p=1)\\
R^{p+1}&(1<p<3\ \wedge\ p\ne2)\\
R^2&(\frac12<p<1)\\
R^2\ln R^{-1}&(p=\frac12)\\
R^{2p+1}&(0<p<\frac12)
\end{cases}.
\end{displaymath}

The limiting behaviour for $R\rightarrow\infty$ on the other hand is
easier to derive because the Fox H-function in equation (\ref{eq:sbf})
for $\mathfrak d>0$ and $s\ne0$ reduces to a power series,
\begin{equation}
\frac{\Sigma(R)}{\bar\rho_\mathrm sr_\mathrm s}=
\biggl(\frac{r_\mathrm s}R\biggr)^{\alpha-1}\frac{\!\sqrt\pi}{X^{\delta-\alpha}}
\sum_{n=0}^\infty\frac{\Gamma\bigl(\frac{\delta-1+pn}2\bigr)}
{\Gamma\bigl(\frac{\delta+pn}2\bigr)}
\frac{(-1)^n\bigl(\frac{\delta-\alpha}p\bigr)_n}{n!X^{pn}},
\end{equation}
which converges for $\abs{X}>1$. Here $X=\delta^{-1/p}(R/r_\mathrm s)$.
This is consistent with equation (\ref{eq:limb}). For the $s=0$ case
however, equation (\ref{eq:limb}) indicates that $\Sigma(R)$ falls
off faster than $R\rho(R)$ and we instead find the following asymptotic
expansion ($R\rightarrow\infty$)
\begin{equation}
\frac{\Sigma(R)}{\!\sqrt{2\pi}R\rho(R)}
\left(\frac R{r_\mathrm s}\right)^{p/2}
\simeq
1-\frac{4\alpha+3p-6}8\left(\frac{r_\mathrm s}R\right)^p+\dotsb.
\end{equation}
That is to say, we find that
$\Sigma(R)\sim\sqrt{2\pi}R^{1-\frac p2}\rho(R)$ for a
3-d density profile behaving as
$\rho(r)\sim Ar^{-\alpha}\rme^{-Br^p}$.

\renewcommand\thetable{\arabic{table}}
\input AZ_fit-tab

\newpage
\section{analytic potential--density pairs}
\label{app:tab}

Thanks to the properties of the beta and gamma functions,
any one of the followings in the list is the sufficient
condition for the potential for our model family
to be written down analytically, namely,
\begin{enumerate}
\item $(\alpha,\delta)=(2-mp,3+np)$ or $(\alpha,\delta)=(3-mp,2+np)$ where
$m$ and $n$ are positive integers
(Tabs.~\ref{tab1} and \ref{tab2}),
\item $p^{-1}$ is a positive integer, and $\tilde\alpha_2/p$ or
$\hat\delta_3/p$ is a positive integer
(Tabs.~\ref{tab3} and \ref{tab4}), or
\item all of $2/p$, $2\alpha/p$, and $2\delta/p$ are integers
(Tabs.~\ref{tab5}-\ref{tab7}).
\end{enumerate}
Tables \ref{tab1} and \ref{tab2} give examples of the case (i), which
include the models studied by \citet{EA05} corresponding to
$(\alpha,\delta)=(2-p,3+p)$. The examples for the case (ii)
are found in Tables \ref{tab3} and \ref{tab4}. These include
the so-called $\gamma$-sphere of \citet{De93} and \citet{Tr94} for which
$\hat\delta_3=p=1$ and the generalized NFW profile studied by \citet{EA06}
for which $\tilde\alpha_2=p=1$. 
This last family is extended to include the
model with $s=0$, i.e., the potential--density pair of
$\psi\propto(\rme^{-\eta r}-1)/r$ and $\rho\propto r^{-1}\rme^{-\eta r}$.
Finally, in Tables \ref{tab5}-\ref{tab7}, we list some examples of
the case (iii) with $p=2$ (Tab.~\ref{tab5}), $p=1$ (Tab.~\ref{tab6}), and
$p=\frac12$ (Tab.~\ref{tab7}), respectively. Some examples
are in fact redundant as they also belong to one or both of the first two
cases. In Table \ref{tab7}, we only list those with both $2\alpha$ and
$2\delta$ are integers, but if both $4\alpha$ and $4\delta$ are
integers with $p=\frac12$, the resulting potential is still analytic.
However, if $2\alpha$ or $2\delta$ is a half-integer in this case, the
resulting expressions involve inverse trigonometric/hyperbolic functions.

The potential generated by the \citeauthor{Ei65} profile with
an integer index $n=p^{-1}$ is another example of the case (ii) that is
expressible using only elementary functions. For the $s=0$ cases,
that $p=n^{-1}$ and $\alpha=2-m/n$ where both $m$ and $n$ are
positive integers is a sufficient condition for the existence of an
expression for the potential written using only elementary
functions, which corresponds to the case (ii). On the other hand, if
both $2/p$ and $2\tilde\alpha_2/p$ are positive integers and $s=0$, the
potential can be written down as an expression involving the error
integral. Some examples are found in Tables \ref{tab5}-\ref{tab7}.


Examination of these tables reveals a symmetry
between the potentials for the models with
$(\alpha,\delta)\leftrightarrow(5-\delta,5-\alpha)$.
This may be explained as follows:
Let us suppose that the potential function $\psi(r)$ is generated by the
double power-law profile so that it is the solution of the
spherical Poisson equation,
\begin{equation}
\frac1{r^2}
\frac\rmd{\rmd r}\!\left\lgroup r^2\frac{\rmd\psi(r)}{\rmd r}\right\rgroup
=-\frac{4\pi G\bar\rho a^\alpha}{r^\alpha(1+x^p)^{(\delta-\alpha)/p}}
\end{equation}
where $\bar\rho a^\alpha=\bar\rho_\mathrm sr_\mathrm s^\alpha$ and $x\equiv r/a$.
On the other hand, the same Laplacian operator
acting on the function $\varphi(r)\equiv r^{-1}\psi(a^2/r)$ yields
\begin{equation}\begin{split}
\frac1{r^2}\frac\rmd{\rmd r}\!\Biggl\lgroup
r^2\frac\rmd{\rmd r}\biggl[\frac1r\psi\Bigl(\frac{a^2}r\Bigr)\biggr]
\Biggr\rgroup
&=\frac{a^4}{r^5}\frac1{w^2}
\frac\rmd{\rmd w}\!\left\lgroup w^2\frac{\rmd\psi(w)}{\rmd w}\right\rgroup
\biggr\rvert_{w=a^2/r}
\\&=-\frac1{a}
\frac{4\pi G\bar\rho a^{5-\delta}}{r^{5-\delta}(1+x^p)^{(\delta-\alpha)/p}}.
\end{split}\end{equation}
In other words, if $\psi(r)$ is the potential corresponding to the double
power-law profile with the parameter set of $(\alpha,\delta,p)$,
then the potential for the parameter set of $(5-\delta,5-\alpha,p)$ is
given by $(r/a)^{-1}\psi(a^2/r)$. This also indicates that
the potential for the model with $\alpha+\delta=5$ (which includes the
hypervirial models of \citealt{EA05} encompassing
the \citeauthor{He90} and the \citeauthor{Sc83}-\citeauthor{Pl11} sphere)
satisfies the relation $a\psi(a^2/r)=r\psi(r)$ for all $r$ with
the properly chosen scale length $a$.

\addtocounter{table}2
\input AZ_fit-tab3
\input AZ_fit-tab4
\input AZ_fit-tab5

\begin{landscape}
\pagestyle{empty}
\voffset=2.6in
\input AZ_fit-tab6
\end{landscape}

\label{lastpage}
\end{document}

%% file: AZ_fit-tab
\begin{table*}
\begin{minipage}{111mm}
\caption{Potential-density pairs for $\alpha=2-mp$ and $\delta=3+np$.}
\label{tab1}
\begin{tabular}{ccll}
\hline
$m$ & $n$ & $C\rho^{-1}$ & $-(4\pi GC)^{-1}\psi$ \\
\hline
1&1&${r^{2-p}(1+x^p)^{2+\frac1p}}$&${\displaystyle
\frac{a^p}{1+p}\frac1{(1+x^p)^{1/p}}
}$\\
1&2&${r^{2-p}(1+x^p)^{3+\frac1p}}$&${\displaystyle
\frac{a^p}{(1+p)(1+2p)}\frac1{(1+x^p)^{1/p}}
\biggl(p+\frac1{1+x^p}\biggr)
}$\\
2&1&${r^{2-2p}(1+x^p)^{3+\frac1p}}$&${\displaystyle
\frac{a^{2p}}{(1+p)(1+2p)}\frac1{(1+x^p)^{1/p}}
\biggl(p+\frac{x^p}{1+x^p}\biggr)
}$\\
1&3&${r^{2-p}(1+x^p)^{4+\frac1p}}$&${\displaystyle
\frac{a^p}{(1+p)(1+2p)(1+3p)}\frac1{(1+x^p)^{1/p}}
\biggl[2p^2+\frac{2p}{1+x^p}+\frac{1+p}{(1+x^p)^2}\biggr]
}$\\
2&2&${r^{2-2p}(1+x^p)^{4+\frac1p}}$&${\displaystyle
\frac{a^{2p}}{(1+2p)(1+3p)}\frac1{(1+x^p)^{1/p}}
\biggl[p+\frac{x^p}{(1+x^p)^2}\biggr]
}$\\
3&1&${r^{2-3p}(1+x^p)^{4+\frac1p}}$&${\displaystyle
\frac{a^{3p}}{(1+p)(1+2p)(1+3p)}\frac1{(1+x^p)^{1/p}}
\biggl[2p^2+\frac{2px^p}{1+x^p}+\frac{(1+p)x^{2p}}{(1+x^p)^2}\biggr]
}$\\
\hline
\end{tabular}
\end{minipage}
\end{table*}
\begin{table*}
\begin{minipage}{103mm}
\caption{Potential-density pairs for $\alpha=3-mp$ and $\delta=2+np$.}
\label{tab2}
\begin{tabular}{ccll}
\hline
$m$ & $n$ & $C\rho^{-1}$ & $-(4\pi GC)^{-1}\psi$ \\
\hline
1&1&${r^{3-p}(1+x^p)^{2-\frac1p}}$&${\displaystyle
\frac{a^p}{1-p}
\Biggl[\frac{(1+x^p)^{1/p}}r-\biggl(\frac1r+\frac1{a}\biggr)\Biggr]
}$\\\
1&2&${r^{3-p}(1+x^p)^{3-\frac1p}}$&${\displaystyle
\frac{a^p}{(1-p)(1-2p)}
\Biggl[\biggl(\frac1{1+x^p}-p\biggr)\frac{(1+x^p)^{1/p}}r
+\frac{p-1}r+\frac p{a}\Biggr]
}$\\
2&1&${r^{3-2p}(1+x^p)^{3-\frac1p}}$&${\displaystyle
\frac{a^{2p}}{(1-p)(1-2p)}
\Biggl[\biggl(\frac{x^p}{1+x^p}-p\biggr)\frac{(1+x^p)^{1/p}}r
+\frac pr+\frac{p-1}{a}\Biggr]
}$\\
2&2&${r^{3-2p}(1+x^p)^{4-\frac1p}}$&${\displaystyle
\frac{a^{2p}}{(1-2p)(1-3p)}
\Biggl\lbrace\biggl[\frac{x^p}{(1+x^p)^2}-p\biggr]\frac{(1+x^p)^{1/p}}r
+\frac pr+\frac p{a}\Biggr\rbrace
}$\\
\hline
\end{tabular}
If the potential results in $\frac00$ overall, it reduces
to an expression involving the logarithmic function by the continuous limit.
See Tabs.~\ref{tab6} (for $p=1$) and \ref{tab7} (for $p=\frac12$).
\end{minipage}
\end{table*}

%% file: AZ_fit-tab3
\begin{table*}
\begin{minipage}{170mm}
\caption{Potential-density pairs for $p=1/n$ and $\delta=3+m/n$.}
\label{tab3}
\begin{tabular}{ccll}
\hline
$\delta$ & $p$ & $C\rho^{-1}$ & $-(4\pi GC)^{-1}\psi$ \\
\hline
4&1&${r^\alpha(1+x)^{4-\alpha}}$&${\displaystyle
\frac{a^{2-\alpha}}{(2-\alpha)(3-\alpha)}
\Biggl[1-\biggl(\frac x{1+x}\biggr)^{2-\alpha}\Biggr]
}$\smallskip\\
5&1&${r^\alpha(1+x)^{5-\alpha}}$&${\displaystyle
\frac{a^{2-\alpha}}{(2-\alpha)(3-\alpha)(4-\alpha)}
\Biggl[2-\frac{2x+4-\alpha}{(1+x)^{3-\alpha}}x^{2-\alpha}\Biggr]
}$\smallskip\\
$\frac72$&$\frac12$&${r^\alpha(1+\!\sqrt x)^{7-2\alpha}}$&${\displaystyle
\frac{4a^{2-\alpha}}{(4-2\alpha)(5-2\alpha)(6-2\alpha)}\Biggl[1-
\frac{\sqrt x+5-2\alpha}{(1+\!\sqrt x)^{5-2\alpha}}x^{2-\alpha}\Biggr]
}$\smallskip\\
4&$\frac12$&${r^\alpha(1+\!\sqrt x)^{2(4-\alpha)}}$&${\displaystyle
\frac{4a^{2-\alpha}}
{(4-2\alpha)(5-2\alpha)(6-2\alpha)(7-2\alpha)}
\Biggl[3-\frac{3x+6(3-\alpha)\!\sqrt x+(5-2\alpha)(7-2\alpha)}
{(1+\!\sqrt x)^{2(3-\alpha)}}x^{2-\alpha}\Biggr]
}$\smallskip\\
$\frac92$&$\frac12$&${r^\alpha(1+\!\sqrt x)^{9-2\alpha}}$&${\displaystyle
\frac{8a^{2-\alpha}}{(4-2\alpha)_5}
\Biggl[6-\frac{6x^{\!\frac32}\!+6(7-2\alpha)x
+(4-\alpha)(29-10\alpha)\!\sqrt x+(4-\alpha)(5-2\alpha)(7-2\alpha)}
{(1+\!\sqrt x)^{7-2\alpha}}x^{2-\alpha}\Biggr]
}$\smallskip\\
5&$\frac12$&${r^\alpha(1+\!\sqrt x)^{2(5-\alpha)}}$&${\displaystyle
\frac{8a^{2-\alpha}}{(4-2\alpha)_6}
\Biggl[30-\frac{30x^2+60(4-\alpha)x^{\!\frac32}\!
+9(9-2\alpha)(10-3\alpha)x
+2(4-\alpha)(9-2\alpha)(20-7\alpha)\!\sqrt x
+8(4-\alpha)\bigl(\frac52-\alpha\bigr)_3}
{(1+\!\sqrt x)^{2(4-\alpha)}}x^{2-\alpha}\Biggr]
}$\\
\hline
\end{tabular}

Here, $(a)_n=(a)_n^+$ is the Pochhammer symbol.
The potentials corresponding to
$(\alpha,p)=(2,1)$, $\bigl(2,\frac12\bigr)$, and
$\bigl(\frac52,\frac12\bigr)$ are found by the continuous limits,
which result in the expressions involving the logarithmic function.
The specific results are provided in
Tabs.~\ref{tab6} ($p=1$) and \ref{tab7} ($p=\frac12$).
\end{minipage}
\end{table*}

%% file: AZ_fit-tab4
\begin{table*}
\begin{minipage}{163mm}
\caption{Potential-density pairs for $p=1/n$ and $\alpha=2-m/n$.}
\label{tab4}
\begin{tabular}{ccll}
\hline
$\alpha$ & $p$ & $C\rho^{-1}$ & $-(4\pi GC)^{-1}\psi$ \\
\hline
1&1&${r\,(1+x)^{\delta-1}}$&${\displaystyle
\frac{a^2}{(\delta-3)(\delta-2)}
\frac1r\Biggl[1-\frac1{(1+x)^{\delta-3}}\Biggr]
}$\smallskip\\
0&1&${(1+x)^\delta}$&${\displaystyle
\frac{a^3}{(\delta-3)(\delta-2)(\delta-1)}
\frac1r\Biggl[2-\frac{2+(\delta-1)x}{(1+x)^{\delta-2}}\Biggr]
}$\smallskip\\
$\frac32$&$\frac12$&${r^\frac32(1+\!\sqrt x)^{2\delta-3}}$&${\displaystyle
\frac{4a^{3/2}}{(2\delta-6)(2\delta-5)(2\delta-4)}\frac1r\Biggl[1-
\frac{1+(2\delta-5)\!\sqrt x}{(1+\!\sqrt x)^{2\delta-5}}\Biggr]
}$\smallskip\\
1&$\frac12$&${r\,(1+\!\sqrt x)^{2(\delta-1)}}$&${\displaystyle
\frac{4a^2}{(2\delta-6)(2\delta-5)(2\delta-4)(2\delta-3)}
\frac1r\Biggl[3-
\frac{3+6(\delta-2)\!\sqrt x+(2\delta-3)(2\delta-5)x}
{(1+\!\sqrt x)^{2(\delta-2)}}\Biggr]
}$\smallskip\\
$\frac12$&$\frac12$&${\sqrt r\,(1+\!\sqrt x)^{2\delta-1}}$&${\displaystyle
\frac{8a^{5/2}}{(2\delta-6)_5}\frac1r
\Biggl[6-\frac{6+6(2\delta-3)\!\sqrt x
+(\delta-1)(10\delta-21)x+(\delta-1)(2\delta-3)(2\delta-5)x^{\!\frac32}}
{(1+\!\sqrt x)^{2\delta-3}}\Biggr]
}$\smallskip\\
0&$\frac12$&${(1+\!\sqrt x)^{2\delta}}$&${\displaystyle
\frac{8a^3}{(2\delta-6)_6}\frac1r
\Biggl[30-\frac{30+60(\delta-1)\!\sqrt x
+9(2\delta-1)(3\delta-5)x
+2(\delta-1)(2\delta-1)(7\delta-15)x^{\!\frac32}\!
+8(\delta-1)\bigl(\delta-\frac52\bigr)_3x^2}
{(1+\!\sqrt x)^{2(\delta-1)}}\Biggr]
}$\\
\hline
\end{tabular}

The potentials corresponding to the zero denominator for $\psi$ result
in a logarithmic expression through the limiting process if $\delta>2$.
The results are given in Tabs.~\ref{tab6} ($p=1$) and
\ref{tab7} ($p=\frac12$). The potential at the limit
of $\delta\rightarrow\infty$ ($s=0$) are
also found in the same tables.
\end{minipage}
\end{table*}

%% file: AZ_fit-tab5
\begin{table*}
\begin{minipage}{3.45in}
\caption{Analytic potentials with $p=2$.
Here, we list $-(4\pi GCa^{2-\alpha})^{-1}\psi$ for $\delta<\infty$ whilst
the corresponding density is given by
$C\rho^{-1}=r^\alpha(1+x^2)^{(\delta-\alpha)/2}$.}
\label{tab5}
\begin{tabular}{clll}
\hline $p=2$ & $\alpha=0$ & $\alpha=1$ & $\alpha=2$
\\\hline
$\delta=3$ &
$\dfrac{\arsinh x}x$ &
$\dfrac{\ln(1+x^2)}{2x}+\arctan\frac1x$ &
$\dfrac{\arsinh x}x+\arsinh\frac1x$
\\
$\delta=4$ &
$\dfrac{\arctan x}{2x}$ &
$\dfrac{1+x-\!\sqrt{1+x^2}}x$ &
$\dfrac{\arctan x}x+\frac12\ln\bigl(1+\frac1{x^2}\bigr)$
\smallskip\\
$\delta=5$ &
$\dfrac{\scriptstyle 1}{3\!\sqrt{1+x^2}}$ &
$\frac12\arctan\frac1x$ &
$\arsinh\frac1x$
\\\hline
$s=0$ &
$\dfrac{\sqrt\pi}{4y}\erf y$ &
$\dfrac{\sqrt\pi}2\erfc y+\dfrac{1-\rme^{-y^2}}{2y}$ &
$\dfrac{\sqrt\pi}{2y}\erf y+\frac12E_1(y^2)$
\\\hline
\end{tabular}

For $s=0$, we list ${-(4\pi GCa^{2-\alpha})^{-1}}\psi$
whose corresponding density is given by
$C\rho^{-1}=r^\alpha\exp(y^2)$. These actually
do not result in potentials that are expressible by elementary
functions, but the expressions involve the error integral or the
exponential $E_1$-integral (for $\alpha=2$).\vfill
\end{minipage}
\end{table*}

%% file: AZ_fit-tab6
\begin{table*}
\begin{minipage}{9.5in}
\caption{Analytic potentials with $p=1$. The normalization is
the same as Tab.~\ref{tab5}.}
\label{tab6}
\begin{tabular}{cllllll}
\hline $p=1$ & $\alpha=0$ & $\alpha=\frac12$ & $\alpha=1$ &
$\alpha=\frac32$ & $\alpha=2$ & $\alpha=\frac52$ \\\hline
$\delta=\frac52$ &
$\frac{16}{3}\left(\dfrac{4+3x}{4\!\sqrt{1+x}}-1\right)\dfrac1x$ &
$3\left(\dfrac1{\sqrt x}-\dfrac{\arctan\!\sqrt x}x\right)+\arctan\dfrac{\scriptstyle1}{\!\sqrt x}$ &
$\dfrac{4\left(\!\sqrt{x+1}-1\right)}x$ &
$2\left(\dfrac1{\sqrt x}-\dfrac{\arctan\!\sqrt x}x+\arctan\dfrac{\scriptstyle1}{\!\sqrt x}\right)$ &
$2\left(\dfrac{\!\sqrt{x+1}-1}x+\arsinh\dfrac{\scriptstyle1}{\!\sqrt x}\right)$ &
$\dfrac4{\sqrt x}$
\smallskip\\
$\delta=3$ &
$\dfrac{\ln(1+x)}x-\dfrac1{2(1+x)}$ &
$\frac23\left(1-\dfrac{x+3}{\sqrt{x\,(x+1)}}\right)+\dfrac{2\arsinh\!\sqrt x}x$ &
$\dfrac{\ln(1+x)}x$ &
$2\left(1-\!\sqrt{\dfrac{1+x}x}+\dfrac{\arsinh\!\sqrt x}x\right)$ &
$\ln\Bigl(1+\dfrac{\scriptstyle1}x\Bigr)+\dfrac{\ln(1+x)}x$ &
$2\left(\!\sqrt{\dfrac{x+1}x}-1+\dfrac{\arsinh\!\sqrt x}x\right)$
\smallskip\\
$\delta=\frac72$ &
$\frac{16}{15}\left[1-\dfrac{4+5x}{4(1+x)^{3/2}}\right]\dfrac1x$ &
$\frac14\left(\arctan\dfrac{\scriptstyle1}{\!\sqrt x}+\dfrac{3\arctan\!\sqrt x}x-\dfrac{x+3}{(x+1)\!\sqrt x}\right)$ &
$\frac43\left(1-\dfrac1{\sqrt{1+x}}\right)\dfrac1x$ &
$\dfrac{\arctan\!\sqrt x}x+\arctan\dfrac{\scriptstyle1}{\!\sqrt x}-\dfrac1{\sqrt x}$ &
$2\left(\dfrac{1-\!\sqrt{1+x}}x+\arsinh\dfrac{\scriptstyle1}{\!\sqrt x}\right)$ &
$2\left(\dfrac1{\sqrt x}+\dfrac{\arctan\!\sqrt x}x-\arctan\dfrac{\scriptstyle1}{\!\sqrt x}\right)$
\smallskip\\
$\delta=4$ &
$\dfrac{1+2x}{6(1+x)^2}$ &
$\frac4{15}\left[1-\left(\dfrac x{1+x}\right)^{\frac32}\right]$ &
$\dfrac1{2(1+x)}$ &
$\frac43\left(1-\!\sqrt{\dfrac x{1+x}}\right)$ &
$\ln\Bigl(1+\dfrac{\scriptstyle1}x\Bigr)$ &
$4\left(\!\sqrt{\dfrac{x+1}x}-1\right)$
\smallskip\\
$\delta=\frac92$ &
$\frac{16}{105}\left[1-\dfrac{4+7x}{4(1+x)^{5/2}}\right]\dfrac1x$ &
$\frac18\left(\arctan\dfrac{\scriptstyle1}{\!\sqrt x}+\dfrac{\arctan\!\sqrt x}x\right)-\dfrac{3x^2+2x+3}{24\!\sqrt x\,(x+1)^2}$ &
$\frac4{15}\left[1-\dfrac1{(1+x)^{3/2}}\right]\dfrac1x$ &
$\frac14\left(\dfrac{\arctan\!\sqrt x}x+3\arctan\dfrac{\scriptstyle1}{\!\sqrt x}-\dfrac{1+3x}{(1+x)\!\sqrt x}\right)$ &
$\frac23\left(1-\dfrac{1+3x}{\sqrt{1+x}}\right)\dfrac1x+2\arsinh\dfrac{\scriptstyle1}{\sqrt x}$ &
$3\left(\dfrac1{\sqrt x}-\arctan\dfrac{\scriptstyle1}{\!\sqrt x}\right)+\dfrac{\arctan\!\sqrt x}x$
\smallskip\\
$\delta=5$ &
$\dfrac{1+3x+x^2}{12(1+x)^3}$ &
$\frac{16}{105}\left[1-\dfrac{(4x+7)\,x^{3/2}}{4(x+1)^{5/2}}\right]$ &
$\dfrac{2+x}{6(1+x)^2}$ &
$\frac{16}{15}\left[1-\dfrac{(4x+5)\!\sqrt x}{4(x+1)^{3/2}}\right]$ &
$\ln\Bigl(1+\frac1x\Bigr)-\dfrac1{2(1+x)}$ &
$\frac{16}3\left[\dfrac{4x+3}{4\!\sqrt{x\,(x+1)}}-1\right]$
\smallskip\\\hline
$s=0$ &
$\dfrac2y\left[1-(1+\frac y2)\,\rme^{-y}\right]$ &
$\dfrac{\sqrt\pi}2\left[1-(1-\frac3{2y})\erf\!\sqrt y\right]-\dfrac{3\rme^{-y}}{2\!\sqrt y}$ &
$\dfrac{1-\rme^{-y}}y$ &
$\sqrt\pi\left[1-(1-\frac1{2y})\erf\!\sqrt y\right]-\dfrac{\rme^{-y}}{\sqrt y}$ & &
\\\hline
\end{tabular}
\end{minipage}
\end{table*}

\begin{table*}
\begin{minipage}{9.5in}
\caption{Analytic potentials with $p=\frac12$. The normalization is
again the same as Tab.~\ref{tab5} and \ref{tab6}.}
\label{tab7}
\begin{tabular}{cllllll}
\hline $p=\frac12$ & $\alpha=0$ & $\alpha=\frac12$ & $\alpha=1$ &
$\alpha=\frac32$ & $\alpha=2$ & $\alpha=\frac52$ \\\hline
$\delta=\frac52$ &
$\dfrac{60+153\!\sqrt x+119x+24x^{\!\frac32}}{6\!\sqrt x\,(1+\!\sqrt x)^3}-\dfrac{10\ell^+}x$ &
$\dfrac{2(12+19\!\sqrt x+6x)}{3\!\sqrt x\,(1+\!\sqrt x)^2}-\dfrac{8\ell^+}x$ &
$\dfrac{2(3+2\!\sqrt x)}{\sqrt x\,(1+\!\sqrt x)}-\dfrac{6\ell^+}x$ &
$\dfrac4{\sqrt x}-\dfrac{4\ell^+}x$ &
$\dfrac2{\sqrt x}+2\ell^--\dfrac{2\ell^+}x$ & $\dfrac4{\sqrt x}$ \smallskip\\
$\delta=3$ &
$\dfrac{2\ell^+}x-\dfrac{60+207\!\sqrt x+248x+107x^{\!\frac32}}{30\!\sqrt x\,(1+\!\sqrt x)^4}$ &
$\dfrac{2\ell^+}x-\dfrac{12+29\!\sqrt x+19x}{6\!\sqrt x\,(1+\!\sqrt x)^3}$ &
$\dfrac{2\ell^+}x-\dfrac{2(3+4\!\sqrt x)}{3\!\sqrt x\,(1+\!\sqrt x)^2}$ &
$\dfrac{2\ell^+}x-\dfrac2{\sqrt x\,(1+\!\sqrt x)}$ &
$2\ell^-+\dfrac{2\ell^+}x-\dfrac2{\sqrt x}$ &
$\dfrac2{\sqrt x}-2\ell^-+\dfrac{2\ell^+}x$ \smallskip\\
$\delta=\frac72$ &
$\dfrac{1+5\!\sqrt x+10x+10x^{\!\frac32}}{30(1+\!\sqrt x)^5}$ &
$\dfrac{1+4\!\sqrt x+6x}{15(1+\!\sqrt x)^4}$ &
$\dfrac{1+3\!\sqrt x}{6(1+\!\sqrt x)^3}$ &
$\dfrac2{3(1+\!\sqrt x)^2}$ &
$2\ell^--\dfrac2{1+\!\sqrt x}$ &
$\dfrac4{\sqrt x}-4\ell^-$ \smallskip\\
$\delta=4$ &
$\dfrac{3+18\!\sqrt x+45x+60x^{\!\frac32}\!+10x^2}{210(1+\!\sqrt x)^6}$ &
$\dfrac{1+5\!\sqrt x+10x+2x^{\!\frac32}}{30(1+\!\sqrt x)^5}$ &
$\dfrac{1+4\!\sqrt x+x}{10(1+\!\sqrt x)^4}$ &
$\dfrac{3+\!\sqrt x}{6(1+\!\sqrt x)^3}$ &
$2\ell^--\dfrac{2(4+3\!\sqrt x)}{3(1+\!\sqrt x)^2}$ &
$\dfrac{2(2+3\!\sqrt x)}{\sqrt x\,(1+\!\sqrt x)}-6\ell^-$ \smallskip\\
$\delta=\frac92$ &
$\dfrac{3+21\!\sqrt x+63x+105x^{\!\frac32}\!+35x^2+5x^{\!\frac52}}{420(1+\!\sqrt x)^7}$ &
$\dfrac{2(1+6\!\sqrt x+15x+6x^{\!\frac32}\!+x^2)}{105(1+\!\sqrt x)^6}$ &
$\dfrac{2+10\!\sqrt x+5x+x^{\!\frac32}}{30(1+\!\sqrt x)^5}$ &
$\dfrac{6+4\!\sqrt x+x}{15(1+\!\sqrt x)^4}$ &
$2\ell^--\dfrac{19+29\!\sqrt x+12x}{6(1+\!\sqrt x)^3}$ &
$\dfrac{2(6+19\!\sqrt x+12x)}{3\!\sqrt x\,(1+\!\sqrt x)^2}-8\ell^-$ \smallskip\\
$\delta=5$ &
$\dfrac{1+8\!\sqrt x+28x+56x^{\!\frac32}\!+28x^2+8x^{\!\frac52}\!+x^3}{252(1+\!\sqrt x)^8}$ & 
$\dfrac{5+35\!\sqrt x+105x+63x^{\!\frac32}\!+21x^2+3x^{\!\frac52}}{420(1+\!\sqrt x)^7}$ &
$\dfrac{10+60\!\sqrt x+45x+18x^{\!\frac32}\!+3x^2}{210(1+\!\sqrt x)^6}$ &
$\dfrac{10+10\!\sqrt x+5x+x^{\!\frac32}}{30(1+\!\sqrt x)^5}$ &
$2\ell^--\dfrac{107+248\!\sqrt x+207x+60x^{\!\frac32}}{30(1+\!\sqrt x)^4}$ &
$\dfrac{24+119\!\sqrt x+153x+60x^{\!\frac32}}{6\!\sqrt x\,(1+\!\sqrt x)^3}-10\ell^-$ \smallskip\\\hline
$s=0$ &
$\dfrac{240}y\biggl[1-\Bigl(1+\!\sqrt y+\frac{9y}{20}+\frac{7y^{\!\frac32}}{60}+\frac{y^2}{60}\Bigr)\,\rme^{-\!\sqrt y}\biggr]$ &
$\dfrac{48}y\biggl[1-\Bigl(1+\!\sqrt y+\frac{5y}{12}+\frac{y^{\!\frac32}}{12}\Bigr)\,\rme^{-\!\sqrt y}\biggr]$ &
$\dfrac{12}y\biggl[1-\Bigl(1+\!\sqrt y+\frac y3\Bigr)\,\rme^{-\!\sqrt y}\biggr]$ &
$\dfrac4y\biggl[1-(1+\!\sqrt y)\,\rme^{-\!\sqrt y}\biggr]$ & &
\\\hline
\end{tabular}
Here, $\ell^+=\ln(1+\sqrt x)$ and
$\ell^-=\ln\bigl(1+\frac1{\sqrt x}\bigr)$.
\label{lastpage}
\end{minipage}
\end{table*}